\documentclass[prb,preprint,titlepage,showpacs,preprintnumbers,amsmath,amssymb,floats]{revtex4-2}
\usepackage{amsmath,mathtools}
\usepackage{natbib}
\usepackage{caption}
\usepackage{subcaption}
\usepackage{float}
\usepackage[pdfencoding=auto,colorlinks=true,linkcolor=blue]{hyperref}
\usepackage{amssymb}
\usepackage{gensymb}
\usepackage{amsthm}
\usepackage{graphicx}
\usepackage{physics}
\usepackage{ragged2e}
\usepackage {xcolor}

\newcommand{\bea}{\begin{eqnarray}}
\newcommand{\eea}{\end{eqnarray}}
\newcommand{\beq}{\begin{equation}}
\newcommand{\eeq}{\end{equation}}

\begin{document}
\title{Pseudogap in electron-doped cuprates: thermal precursor to magnetism}
\author{E. K. Kokkinis}
\affiliation{School of Physics and Astronomy and William I. Fine Theoretical Physics Institute, University of Minnesota, Minneapolis, MN 55455, USA}
\author{A. V. Chubukov}
\affiliation{School of Physics and Astronomy and William I. Fine Theoretical Physics Institute, University of Minnesota, Minneapolis, MN 55455, USA}
\date{\today}
\begin{abstract}
We study pseudogap behavior in a metal near an antiferromagnetic instability and apply the results to electron-doped cuprates.  We associate pseudogap behavior with thermal magnetic fluctuations and compute the fermionic self-energy along the Fermi surface in a weak pseudogap regime, which we justify. We analyze the spectral function as a function of frequency (energy distribution curves, EDC) and momentum (momentum distribution curves, MDC).   We show that the EDC  display pseudogap behavior with peaks at a finite frequency at all momenta.  On the other hand, MDC peaks disperse within the pseudogap, ending at a gossamer Fermi surface.  We analyze magnetically-mediated superconductivity and show that in a weak pseudogap regime thermal fluctuations almost cancel out in the gap equation. We favorably compare our results with recent ARPES study [K-J Xu et al,  Nat. Phys. \textbf{19}, 1834–1840 (2023)].
\end{abstract}
\maketitle
\section{Introduction}

The origin of pseudogap behavior observed in cuprates and other  materials ~\cite{Norman_review,Taillefer_review,*Taillefer_review_2,NormalPseudogapReivew2005,
review,Damascelli_review,Vishik_review,Campuzano_review,Kordyuk_review,Armitage_review,Greene_review,Meng2009} remains one of the most challenging problems in the physics of strongly correlated electron systems. The existing theories of the pseudogap can be broadly divided into two categories.  In the first, pseudogap is viewed as a state with some order, either a conventional one, bilinear in fermions (e.g., a charge-density-wave~\cite{Fradkin_2015,*Metlitski_2010,*WangYuxuan2014}, loop current~\cite{Varma_2014} or a pair density wave~\cite{Lee_2014,*Agterberg_2020,*WangYuxuan2015prl}), or a  topological one, which breaks Luttinger theorem without a conventional order~\cite{Rice_2006,Phillips2020,Fabrizzio_2022,*Fabrizzio_2023,Christos_2023,Stepanov_2024}. Within this scenario, pseudogap behavior is the result of order-induced reconstruction of the Fermi surface~\cite{Christos_2024}. In the second category, pseudogap is viewed as a precursor to either superconductivity~\cite{Randeria1998,imp_cuprates3,Berg_2007,Dai_2021,paper_5} or density-wave order, like antiferromagnetism~\cite{Vilk1996,*Vilk1997,*Senechal_2004,Schmalian1999,Sadovskii_review,Moca2000,Sedrakyan2010,Gunnarsson2015,*Gull2015,*Schafer2021,Krien_2021,Berg_2023,Ye_2023,*Ye_2023_1,Simkovic_2024} or pair-density-wave~\cite{Dai_2020}. In this scenario, the Fermi surface is not reconstructed and the spectral weight is finite at all Fermi momenta,  but a portion of the spectral weight is transformed away from the lowest energies, creating a depletion of the density of states around the Fermi energy.

Theories from both categories have been applied to hole-doped cuprates. There is experimental evidence in favor of a true reconstruction of the Fermi surface (see e,g., \cite{Sachdev_2025} and references therein), but some data like Fermi arcs  can be interpreted as evidence in favor of a reduction of the spectral weight near the corners of the Brillouin zone
 without Fermi surface reconstruction (see e,g., \cite{abanov_2000fingerprints,Chubukov2007}).  The data analysis in hole-doped cuprates is further complicated by the presence of strong charge-density-wave correlations, stripes, which replace a commensurate $(\pi,\pi)$ magnetism~\cite{Tranquada_2020,Wietek_2021}, and $B_{1g}$ phonons~\cite{Devereaux_2004}.

The situation looks less complicated in electron-doped (n-type) cuprates, like Nd$_{2-x}$Ce$_x$CuO$_4$ (NCCO).  Here, a plethora of experimental data  shows~\cite{Armitage_review,Greene_review,Motoyama_2007,Matsuda_1992} that a  magnetic order is commensurate, with momentum  ${\bf Q} = (\pi,\pi)$, and it holds up to much larger dopings than for p-type cuprates. This also agrees with previous theoretical analyses \cite{Senechal_2004,Weber_2010,Kyung_2004}. Charge-density-wave fluctuations and effects from phonons are weaker, so it is likely that $(\pi,\pi)$ magnetic fluctuations  play the  major role. At small $x$, the data in the $(\pi,\pi)$ ordered state show~\cite{Armitage_review,Armitage_2001_1,Armitage_2001_2} that electronic excitations are reconstructed into electron pockets around $(\pi,0)$ and $(0,\pi)$. In between,  fermionic excitations are gapped.   At $x =0.15$, which is close to the end point of an antiferromagnetic order, angle resolved photoemission (ARPES) data show superconductivity at $T \approx 25K$ and pseudogap behavior at higher temperatures~\cite{Xu_2023,Matsui_2005_2,Sato_2001,Horio_2019}, i.e., superconductivity emerges out of a pseudogap state.

From theory perspective, the dominance of $(\pi,\pi)$  magnetic fluctuations still allows two scenarios for the pseudogap: a topological order, whose effect on the electrons mimics that from a true $(\pi,\pi)$ order and gives rise to Fermi surface reconstruction into small electron and hole  pockets~~\cite{Christos_2024}, and a precursor to antiferromagnetism~\cite{Vilk1996,*Vilk1997,*Senechal_2004,Schmalian1999,Sadovskii_review,Moca2000,Sedrakyan2010,Gunnarsson2015,*Gull2015,*Schafer2021,Krien_2021,Berg_2023,Ye_2023,*Ye_2023_1,Simkovic_2024}.

In this communication we consider the precursor scenario. We conjecture that pseudogap emerges due to {\it thermal} antiferromagnetic fluctuations - the ones that destroy long-range antiferromagnetic order in quasi-2D systems at temperature $T_N$ well below the exchange energy $J$, and consider how thermal magnetic fluctuations above $T_N$ affect fermionic spectral function, and how this affects superconductivity. We perform controlled perturbative calculations in a weak pseudogap regime, where the pseudogap is smaller than $v_F \xi^{-1} (T)$, where $\xi (T)$ is a magnetic correlation length.

One of our goals is to understand the results of the recent extensive ARPES study of NCCO at $x =0.15$ by Xu et al~\cite{Xu_2023} (see also \cite{Xu_2024}). The authors of~\cite{Xu_2023}  collected ARPES data along several  cuts parallel to the Brillouin zone diagonal (see Fig. \ref{fig:1}) and  extracted from the data the fermionic spectral function  $A(k, \omega)$ as a function of frequency  at a fixed momentum (Energy Distribution Curves (EDC)) and as a function of momentum at a fixed frequency (Momentum Distribution Curves (MDC)). This study found two fundamental features. First, above the superconducting $T_c$, data show that the maximum of the EDC spectra remains at a finite frequency at all momenta $k$, including those where the Fermi surface would be located in the absence of antiferromagnetic fluctuations (we label these momenta as $k_F$). Xu et al interpreted this result as evidence for Fermi surface reconstruction without a $(\pi,\pi)$ order. At the same time, the peak in the MDC data in each cut disperses inside the EDC gap and at  vanishing frequency is located very close to $k_F$. This is what one would expect in a metal with an unreconstructed Fermi surface (gossamer Fermi surface, in the terminology of Xu et al). Second, below $T_c$, Xu et al found a $d-$wave superconducting gap, with a maximum  at the hot spot ${\bf k} = {\bf k}_{h}$ (one of 8 Fermi surface points for which ${\bf k}_{h} + {\bf Q}$ is also on the Fermi surface), where the suppression of the low-energy spectral weight in the EDC data is the largest. This result is seemingly against a common sense argument that in a one-band system gap variation should generally follow variation of the density of states (Xu et al called it a paradox).

We argue that both sets of results are reproduced within the precursor scenario based on thermal antiferromagnetic fluctuations~\cite{Schmalian1999,Sedrakyan2010,Ye_2023,*Ye_2023}. We recall that within this scenario, Fermi surface is {\it not reconstructed}, but some portion of the  spectral weight gets transferred to a finite frequency.   We show that in the magnetic pseudogap phase, the EDC spectral function $A_k (\omega)$  has a maximum at a finite frequency, which is the largest at $k_F$ in each cut. This gives rise to re-entrant behavior of the dispersion extracted from EDC, when ${\bf k}$ varies through ${\bf k}_F$.  The largest density of states depletion among all cuts is at a hot spot ${\bf k}_h$. Both results are consistent with the data in ~\cite{Xu_2023}. Next, we show that MDC spectral function $A_\omega (k)$   taken at a frequency within the gap, has a peak, and the peak position disperses towards $k_F$ as $\omega$ gets smaller. This is again consistent with ~\cite{Xu_2023}.  Finally, we  analyze  superconductivity out of a pseudogap state  and argue that there is no paradox. Namely, we  show that thermal fluctuations almost cancel out in the gap equation in a manner similar to cancellation of non-magnetic impurities. This effect has been discussed before within the Eliashberg theory which assumes a factorization of momentum integration between bosons and fermions~\cite{Millis_1988}. We show  that it holds if we compute the one-loop self-energy exactly,  without factorization. As a result, thermal depletion of the low-energy spectral weight near a hot spot does not affect the angular dependence of the superconducting gap), which remains nearly the same as if  only quantum $(\pi, \pi)$ fluctuations were present.  In the last case, at  weak and moderate couplings, superconducting gap has the largest value at a hot spot~\cite{Abanov_2008}.  This is fully consistent with the data in ~\cite{Xu_2023}.

\section{Thermal antiferromagnetic fluctuations}
The input for our studies is the work by Ye et al \cite{Ye_2023,Ye_2023_1}, which in turn is based on earlier works by Vilk, Tremblay and Senechal~\cite{Vilk1996,*Vilk1997,*Senechal_2004}. Ye et al considered fermions with $t,t'$ dispersion, bandwidth $W = 8t$ and Hubbard interaction $U \geq W$,  and  analyzed the effect of thermal antiferromagnetic fluctuations for a fermion  at a hot spot in a temperature range  between a low $T < T_N$, when long-range $(\pi,\pi)$ order is present in a quasi-2D material, and  higher $T > T_N$, when magnetic order disappears.   They argued that there are two distinct pseudogap regimes at $T > T_N$, which they identified as strong pseudogap and weak pseudogap.  In the strong pseudogap regime, the pseudogap scale $\Delta_{PG}$ is set by Mott physics and is approximately $U$. Simultaneously, the chemical potential strongly shifts from its free-Fermion value $\mu_0 \sim t'$ to $\mu =-U/2 + O(J)$ ($J =4t^2/U$), such that both EDC and MDC ARPES intensities (the products of the spectral functions  and the Fermi function $n_F (\omega)$) have peaks at $\omega \sim -J$.  This regime holds at smaller dopings and at $T \geq T_N$,  when $\Delta_{PG}$  is larger than $v_F \xi^{-1} (T)$, where $v_F$ is the Fermi velocity and $\xi (T)$ is a magnetic correlation length. To describe this regime diagrammatically, one needs to sum up an infinite series of terms for the fermionic self-energy going beyond self-consistent one-loop, and an infinite set of terms for the magnetic polarization~\cite{Schmalian1999,Sadovskii_review,Sedrakyan2010,Ye_2023,*Ye_2023}. Weak pseudogap regime holds at larger dopings/larger $T$,  when the strength of a magnetic order at $T=0$ is reduced and there emerges at wide range of $T > T_N$ where the correlation length $\xi (T)$ remains large, but  $\Delta_{PG}$ becomes smaller than $v_F \xi^{-1} (T)$. In this regime, the  chemical potential $\mu$ remains close to bare $\mu_0$  and the thermal self-energy can be computed perturbatively in the dimensionless coupling $\lambda_{th} = 3 {\bar g} T/(2\pi (v_F \xi^{-1} (T))^2)$, where ${\bar g} \sim U^2/J$ is an effective spin-fermion vertex. A weak pseudogap behavior emerges when $\lambda_{th}$ exceeds a threshold $\lambda_{th,c}\approx0.5$, but is not large (this has to be verified in explicit calculations at a given $\lambda_{th} > \lambda_{th,c}$. Strong pseudogap behavior in turn holds for large $\lambda_{th}$
\footnote{In the strong coupling regime at small $x$,  $T_N \sim J$, hence $T >T_N$ is also of order $J$. Then, even when $\xi (T)$ reduces to a few interatomic spaces and $v_F \xi^{-1}$ becomes of order $t$, $\lambda_{th} \sim {\bar g} T/(v_F \xi^{-1})^2  \sim (U/t)^2 (T/J)$ remains large, of order $U/J$.}. 
For dopings  $x \leq x_c$ (the end point of an antiferromagnetic order), strong pseudogap behavior holds only in a narrow $T$ range immediately above $T_N (x)$, while the bulk of the pseudogap range falls into a weak pseudogap regime.

\begin{figure}[H]
\begin{center}
\includegraphics[width=.43\textwidth]{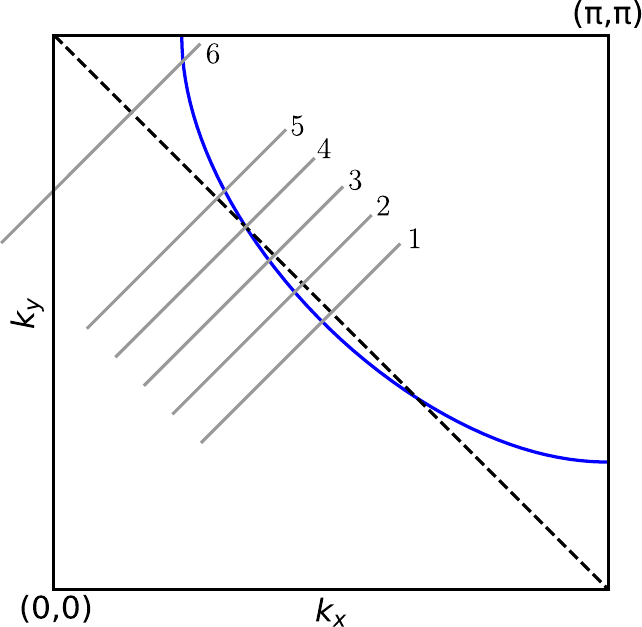}
\end{center}
\caption{
Free fermion dispersion of $t-t'$ model for $t=0.39eV$, $t'=-0.09eV$ and $\mu=-0.08eV$. We study the spectral function along 6 distinct momentum cuts indicated by the gray lines. The black dotted line represents the magnetic Brillouin zone boundary.}
\label{fig:1}
\end{figure}
\newpage
The data from \cite{Xu_2023} for NCCO at $x =0.15$ show that (i) the pseudogap energy $\Delta_{PG}$ is below $100 meV$, smaller than $v_F \xi^{-1} (T) \geq 100 meV$, and (ii)  the gossamer Fermi surface crosses the magnetic Brillouin zone boundary approximately at the same $k_h$ as for free fermions.   The combination of these two facts strongly indicates that the data is obtained in the weak pseudogap regime. We assume that this is the case and compute EDC and MDC spectral functions within one-loop approximation along six cuts parallel to the Brillouin zone diagonal shown in Fig. \ref{fig:1}. Specifically, we compute the thermal self-energy $\Sigma_{th} (k, \omega)$  as the convolution of the static antiferromagnetic  propagator $\chi (q+Q) \propto 1/(q^2 + \xi^{-2} (T))$ and  a propagator of a free fermion $G_0(k+q +Q, \omega)$. We emphasize that we must use the bare fermionic Green's function rather than the full one (i.e., use perturbative rather than self-consistent one-loop approximation), because for thermal fluctuations self-energy and vertex correction insertions have to be treated on equal footings. In the weak pseudogap regime, {\it both} corrections can be neglected, in the strong pseudogap regime, {\it both} have to be included~\cite{Vilk1996,*Vilk1997,*Senechal_2004,Fujimoto_2002,*Yanase_2004,Schmalian1999,Sadovskii_review,Ye_2023}. We explicitly verified (see Appendix \ref{AppendixB}) that in our case ($\lambda_{th} = 0.85$) the two-loop diagrams with self-energy insertion into the propagator of an intermediate fermion (a member of self-consistent one-loop series) and with a vertex correction almost cancel each other, i.e., a perturbative one-loop approximation is quite accurate.

\section{Spectral function in the normal state}

We set $T$ to be fixed, treat $\xi (T)$ as an input parameter, and  measure $\omega$, the dispersion $\epsilon^*_k = \epsilon_k - \mu$, $\Sigma (k, \omega)$  and $G(k, \omega) = (\omega - \epsilon^{*}_k  - \Sigma (k, \omega))^{-1}$  in units of $v_F \xi^{-1}$ and $q$ in units of $\xi^{-1}$. The one-loop expression for the self-energy is~\cite{Vilk_1997,Ye_2023}

\bea
 \Sigma_{th} (k, \omega) &=& \frac{3 {\bar g} T}{(v_F \xi^{-1})^2} \int \frac{d^2 q}{(2\pi)^2} \frac{1}{\omega - \epsilon^*_{k+Q} - q_{\perp}} \frac{1}{q^2_{\perp} + q^2_{\parallel}+ 1} \nonumber \\
 &=& \lambda_{th}\left[ \frac{\log{\left(\omega - \epsilon^{*}_{k+Q} + \sqrt{1+\left(\omega - \epsilon^{*}_{k+Q}\right)^2}\right)}}{\sqrt{1 + \left(\omega - \epsilon^{*}_{k+Q}\right)^2}}
 -i \frac{\pi}{2 \sqrt{1 + \left(\omega - \epsilon^{*}_{k+Q}\right)^2}}\right]
\label{1}
 \eea
\newpage
Substituting this self-energy into $G(k, \omega)$, we obtain after a simple algebra the  spectral function $A(k, \omega) = (1/\pi) |\text{Im} G(k, \omega)|$ in the form
 \beq
  A (k, \omega) = \frac{\lambda_{th}}{2} \frac{\sqrt{1 + \left(\omega - \epsilon^{*}_{k+Q}\right)^2}}{\frac{\pi^2 \lambda_{th}^2}{4} + \left(\lambda_{th} \log{\left(\omega - \epsilon^{*}_{k+Q} + \sqrt{1+ \left(\omega - \epsilon^{*}_{k+Q}\right)^2}\right)}- (\omega - \epsilon^{*}_k)\sqrt{1 + \left(\omega - \epsilon^{*}_{k+Q}\right)^2}\right)^2}
  \label{2}
 \eeq

\begin{figure}[H]
\centering
\begin{minipage}{.32\textwidth}
    \centering
    \includegraphics[width=\linewidth,height=4cm]{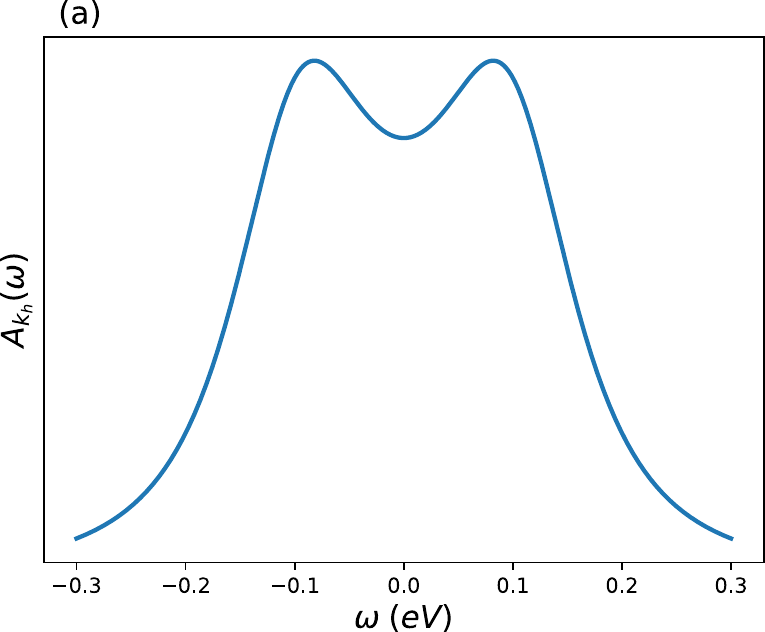}
    \phantomsubcaption\label{fig:2a}
\end{minipage}
\hfill
\begin{minipage}{.32\textwidth}
    \centering
    \includegraphics[width=\linewidth,height=4cm]{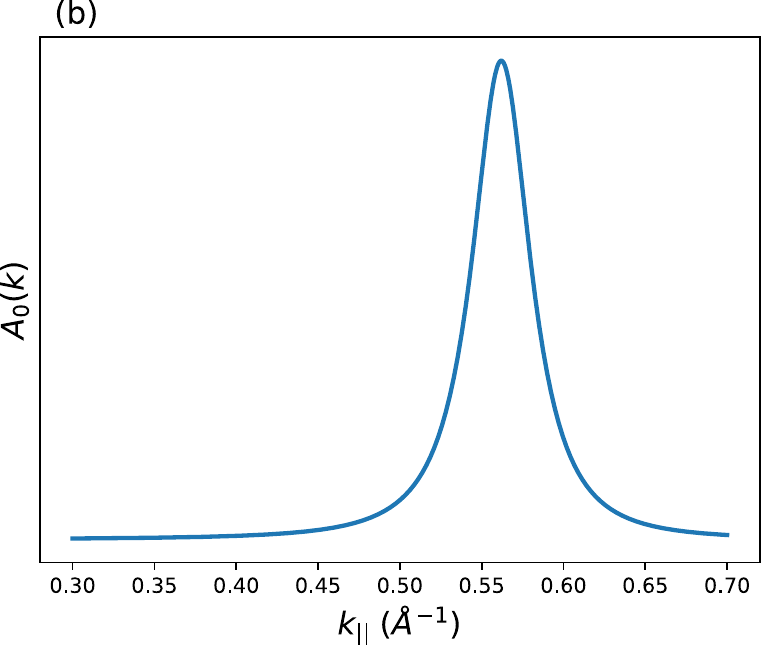}
    \phantomsubcaption\label{fig:2b}
\end{minipage}
\hfill
\begin{minipage}{.32\textwidth}
    \centering
    \includegraphics[width=\linewidth,height=4cm]{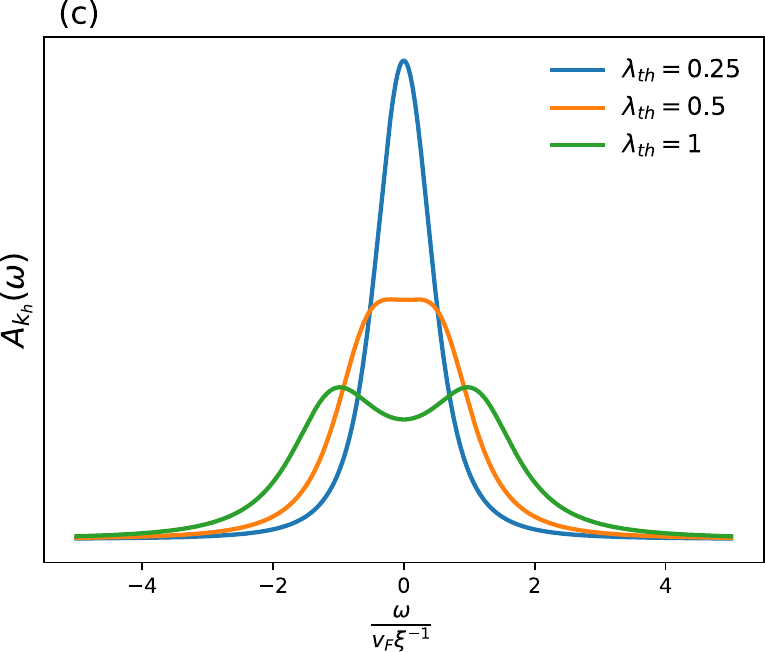}
    \phantomsubcaption\label{fig:2c}
\end{minipage}
\caption{(a)  The spectral function at the hot spot $k=k_h$ as a function of frequency $\omega$. (b) The spectral function at $\omega=0$ plotted along a diagonal momentum cut 4, passing through the hot spot. (c) The spectral function at the hot spot $k=k_h$ as a function of frequency $\omega$ for various different values of the thermal coupling $\lambda_{th}$. Pseudogap peaks appear at $\lambda_{th}>\lambda_{th,c}\sim0.5$. We present the real and imaginary parts of the one-loop self-energy in Appendix \ref{AppendixB}. }
\label{fig:2}
\end{figure}

For practical calculations  we will be using experimentally motivated  parameters $t=0.39eV$, $t'=-0.09eV$, $\mu=-0.08eV$,  $a=3.95\AA$ and $\xi=10a$ (see e.g. Ref. \cite{Motoyama_2007}) and for definiteness set $\lambda_{th}=0.85 > \lambda_{th,c}$. We plot EDC $A_{k_h} (\omega)$ and MDC $A_0 (k)$ along cut 4 in Figs. \ref{fig:2a},\ref{fig:2b}.  We see that despite the apparent similarity between the dependence on momentum and on frequency in Eq. (\ref{2}),  the two functions differ qualitatively.  In particular,  EDC $A_{k_h} (\omega)$ at a hot spot has two peaks at a finite $\omega = \pm \Delta_{PG}\propto (\lambda_{th}-\lambda_{th,c})$, (Fig. \ref{fig:2a}), while MDC $A_{0} (k)$ has a single peak at $k = k_F$ (Fig. \ref{fig:2b}).  The difference comes from the fact that $\epsilon^{*}_k$ and $\epsilon^{*}_{k+Q}$ have opposite signs for the same $k$ in cut 4.    Approximating $\epsilon^{*}_{k+Q}$ as $-\alpha \epsilon^{*}_k = -\alpha v_F (k-k_h)$ and expanding in Eq. (\ref{2}) near $\omega=0$ and $k=k_h$, we obtain
 \bea
&& A_{k_h} (\omega) = \frac{\pi \lambda_{th} }{2} \frac{1}{\lambda_{th}^2 \pi^2/4 + \omega^2 \left((1-\lambda_{th})^2 - \pi^2\lambda_{th}^2/8\right)} \nonumber \\
&&A_{0} (k) = \frac{\pi \lambda_{th} }{2} \frac{1}{\lambda_{th}^2 \pi^2/4 + v^2_F (k-k_h)^2  \left((1+\alpha \lambda_{th})^2 - \pi^2 \alpha^2 \lambda_{th}^2/8\right)}
\label{3}
\eea
We see that pseudogap in the EDC $A_{k_h} (\omega)$ (sign change of the prefactor for $\omega^2$ term) appears at $\lambda_{th} > \lambda_{th,c} = 1/\left(1 + \pi/\left(2\sqrt{2}\right)\right) \approx 0.47$ (Fig. \ref{fig:2c}), while MDC $A_{0} (k)$ remains peaked at $k=k_h$ for all $\lambda_{th} > 0.47$  for which the prefactor for $(k-k_k)^2$ term remains positive
\footnote{For a more accurate treatment, one should keep the quantum self-energy. The latter is  well described within Eliashberg theory and predominantly depends on frequency, $\Sigma_{qm} \approx \Sigma_{qm} (\omega)$ (see, e.g., Ref. \protect\cite{Abanov_2003}). At small frequencies, $\omega - \Sigma_{qm} (\omega) \approx \omega(1+\lambda_{qm})$. This renormalization  rescales frequency, but other than this the results remain the same as without $\Sigma_{qm} (\omega)$. We note in this regard that even at strong coupling, quantum self-energy for a dispersing boson (our case) gives rise to an incoherent non-Fermi liquid behavior, but does not give rise to pseudogap behavior.}.
We re-iterate that to obtain $\lambda_{th,c}$ and the pseudogap behavior at $\lambda_{th} > \lambda_{th,c}$, we need to use perturbative rather than self-consistent one-loop approximation. We discuss $\Sigma_{th} (k, \omega)$ in more detail in Appendices \ref{AppendixA} and \ref{AppendixB}. In Appendix \ref{AppendixA}, we obtain $\Sigma_{th} (k, \omega)$ analytically in a self-consistent one-loop approximation and show that then the spectral function does not display a pseudogap behavior at any $\lambda_{th}$. In Appendix \ref{AppendixB} we compute $\Sigma_{th} (k, \omega)$ at two-loop order by combining the contributions from the diagram with the self-energy of an intermediate fermion (part of diagrammatic series for a self-consistent one-loop) and the diagram with the vertex correction. We show that the two almost cancel each other at small frequencies and hence have almost zero effect on the value of $\lambda_{th,c}$. For completeness, in Appendix \ref{AppendixC} we discuss full series for the thermal self-energy and show that vertex correction diagrams are essential for the description of the pseudogap behavior. We note in passing that  a double peak structure of $A_{k_{h}}(\omega)$ in EDC due to  thermal fluctuations has been obtained at moderate coupling using the functional Renormalization Group method \cite{Rohe_2005}.

\begin{figure}[H]
\centering
\begin{minipage}{.32\textwidth}
    \centering
    \includegraphics[width=\linewidth,height=4cm]{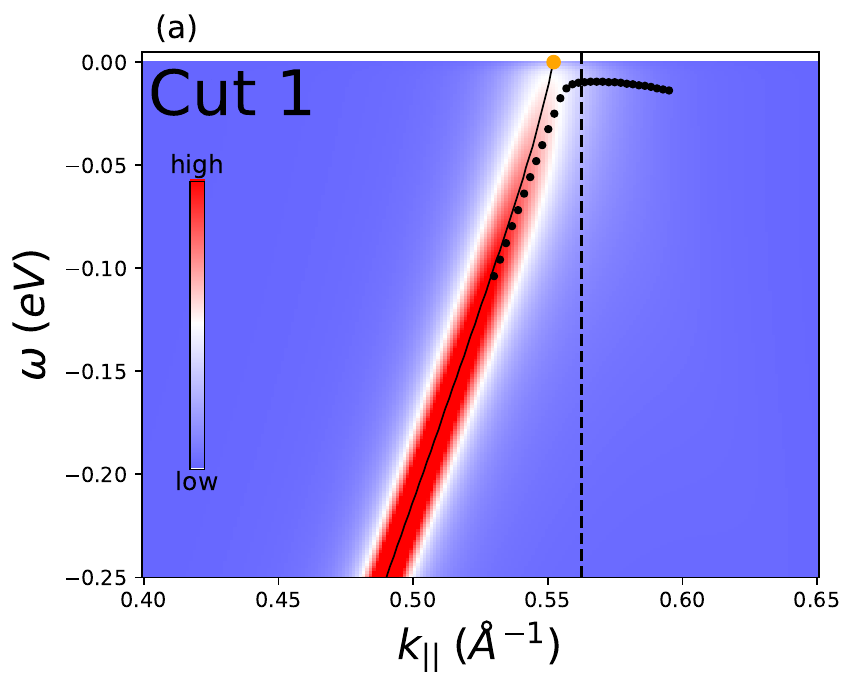}
    \phantomsubcaption\label{fig:3a}
\end{minipage}
\hfill
\begin{minipage}{.32\textwidth}
    \centering
    \includegraphics[width=\linewidth,height=4cm]{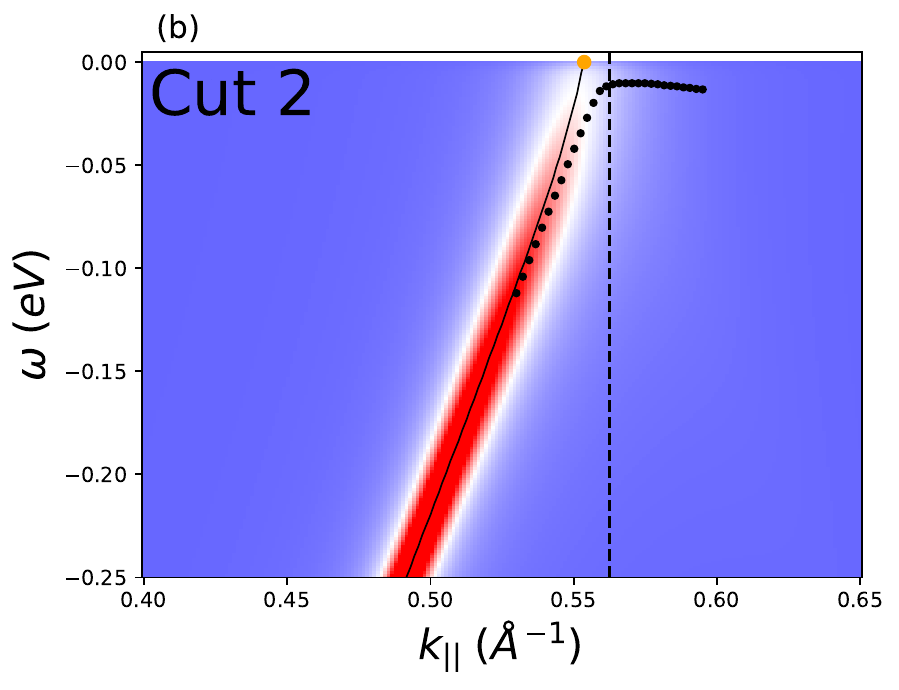}
    \phantomsubcaption\label{fig:3b}
\end{minipage}
\hfill
\begin{minipage}{.32\textwidth}
    \centering
    \includegraphics[width=\linewidth,height=4cm]{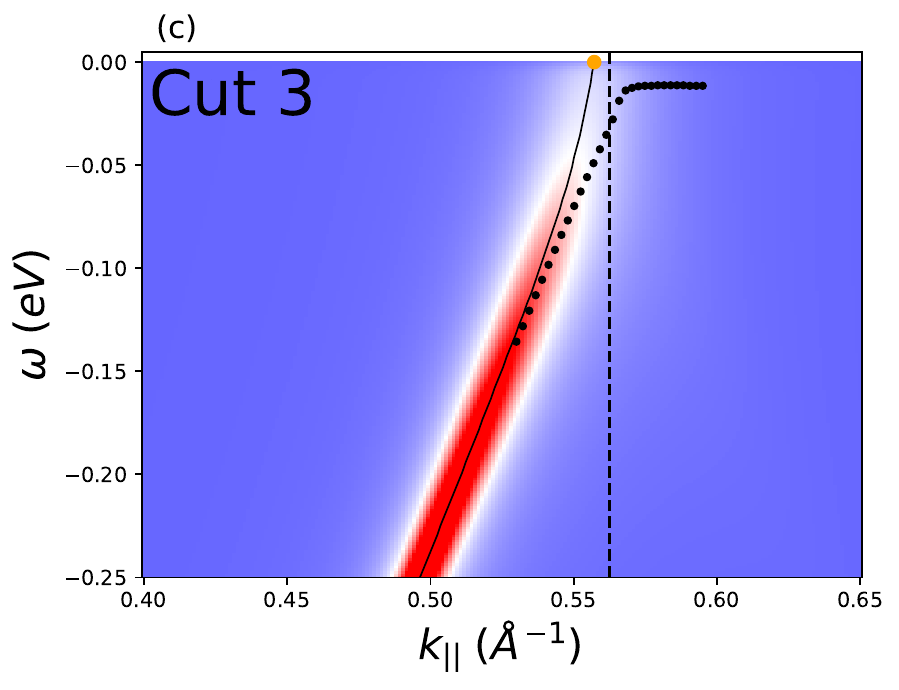}
    \phantomsubcaption\label{fig:3c}
\end{minipage}
 \vspace{-1cm}
\begin{minipage}{.32\textwidth}
    \centering
    \includegraphics[width=\linewidth,height=4cm]{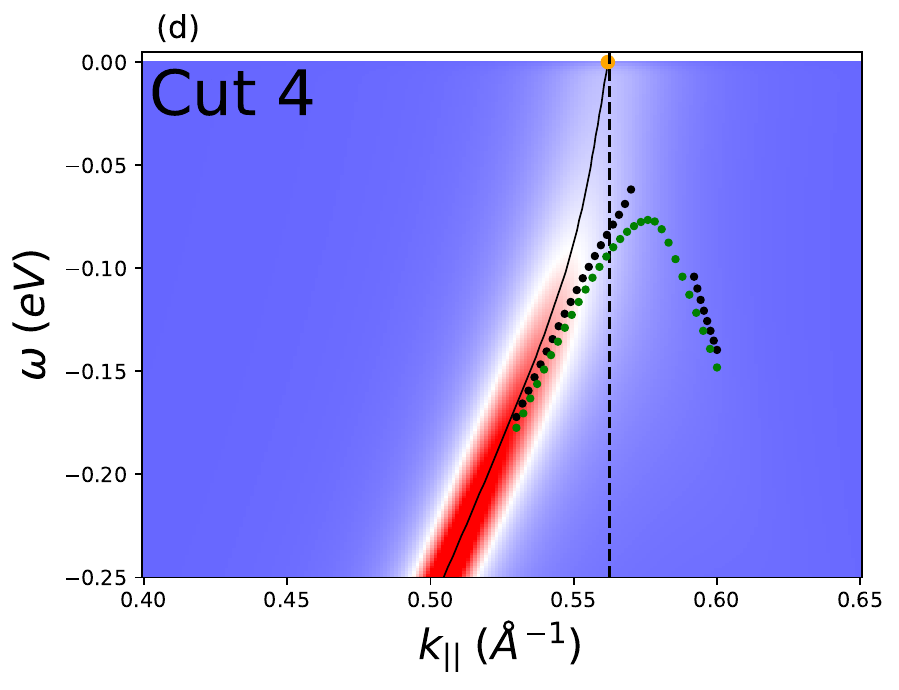}
    \phantomsubcaption\label{fig:3d}
\end{minipage}
\hfill
\begin{minipage}{.32\textwidth}
    \centering
    \includegraphics[width=\linewidth,height=4cm]{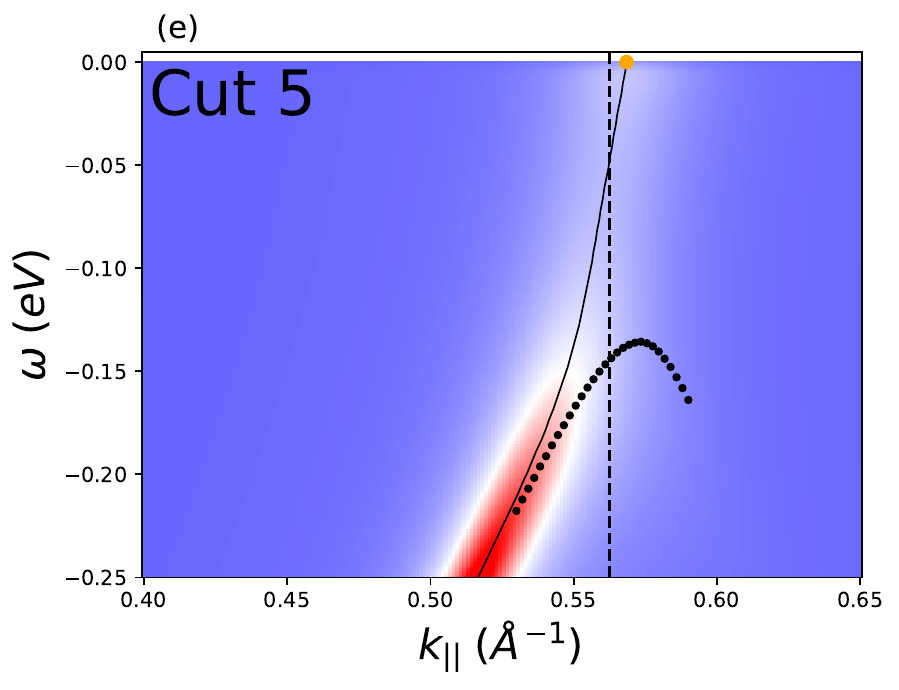}
    \phantomsubcaption\label{fig:3e}
\end{minipage}
\hfill
\begin{minipage}{.32\textwidth}
    \centering
    \includegraphics[width=\linewidth,height=4cm]{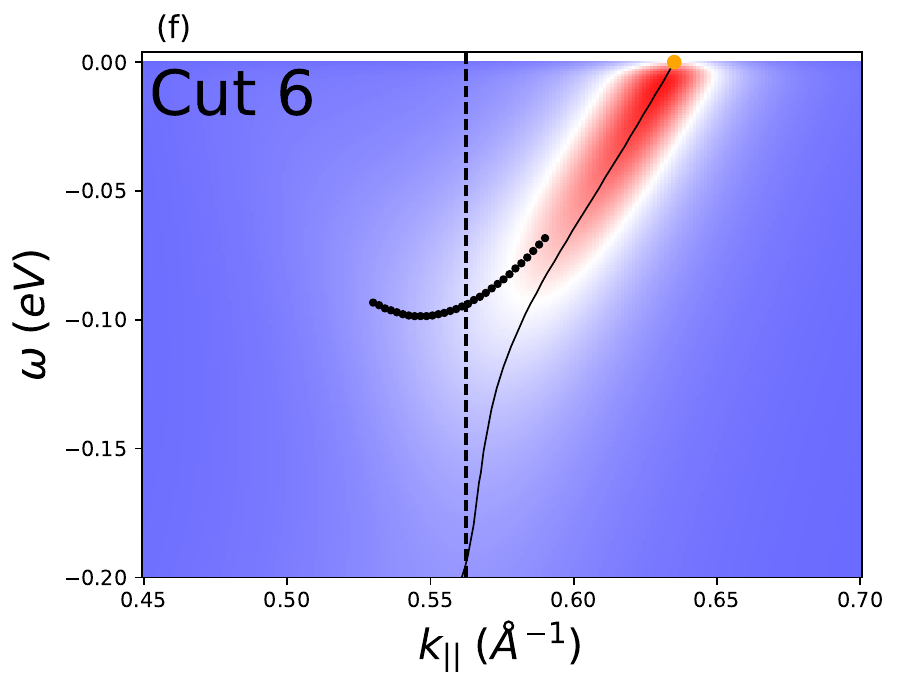}
    \phantomsubcaption\label{fig:3f}
\end{minipage}
\caption{(a)-(f)  Frequency-Momentum Spectra for cuts 1-6. The black dots correspond to the positions of the EDC peaks for different values of momentum. The black line tracks the position of the MDC peak at different frequencies, with the orange dot representing $k_F$. The dashed line represents the magnetic Brillouin zone boundary. The green dots in (d) correspond to the position of the EDC peaks for a larger value of the thermal coupling $\lambda_{th}=0.95$.}
\label{fig:3}
\end{figure}

In Figs. \ref{fig:3a}-\ref{fig:3f} we plot EDC and MDC spectral intensities $I_{EDC}$ and $I_{MDC}$   for finite frequencies and momenta along the same cuts as in Ref. \cite{Xu_2023} (shown in Fig. \ref{fig:1}). We see that $I_{EDC}$ clearly displays a pseudogap behavior, i.e., peak position remains at a finite frequency at $k=k_F$, while the peak in $I_{MDC}$ closely follows the dispersion of a free fermion $\omega = \epsilon^*_k$ and at $\omega =0$ determines the  Fermi surface of free fermions. This is a gossamer Fermi surface in the notations of \cite{Xu_2023}. In addition, we reproduce the non-monotonic evolution of the EDC peak position that is observed experimentally. We emphasize that this is obtained in the absence of Fermi surface reconstruction. Note that in cut 4, the negative energy EDC peak disappears for certain momentum values. This is consistent with the experimental observation in ~\cite{Xu_2023})  that the EDC peaks in this momentum range are not well defined. In theoretical EDC, this occurs because for those momenta the critical thermal coupling $\lambda_{th,c}(k)$, required for a two-peak structure in EDC, exceeds our $\lambda_{th}=0.85$. As we move further along the cut and $\lambda_{th,c}(k)$ decreases, the pseudogap peak gradually re-emerges in the spectral intensity. For  larger  $\lambda_{th}$ ($\lambda_{th} =0.95$), we did find that the EDC peaks  are observable for all values of momentum (green dots in Fig. \ref{fig:3d}).  The analysis of EDC and MDC for cut 6 requires a more detailed discussion. We present it in the Appendix \ref{AppendixD}.

\begin{figure}[H]
\centering
\begin{minipage}{.32\textwidth}
    \centering
    \includegraphics[width=\linewidth]{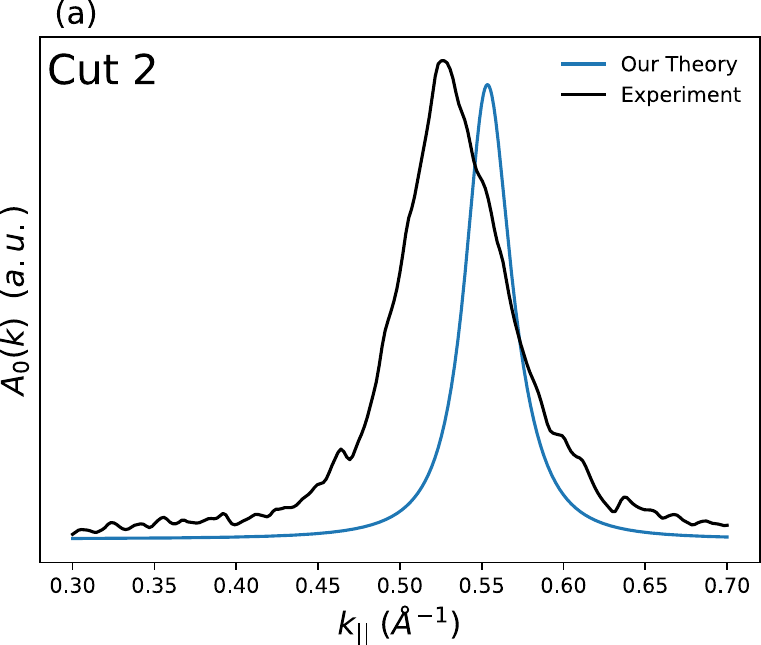}
    \phantomsubcaption\label{fig:4a}
\end{minipage}
\hfill
\begin{minipage}{.32\textwidth}
    \centering
    \includegraphics[width=\linewidth]{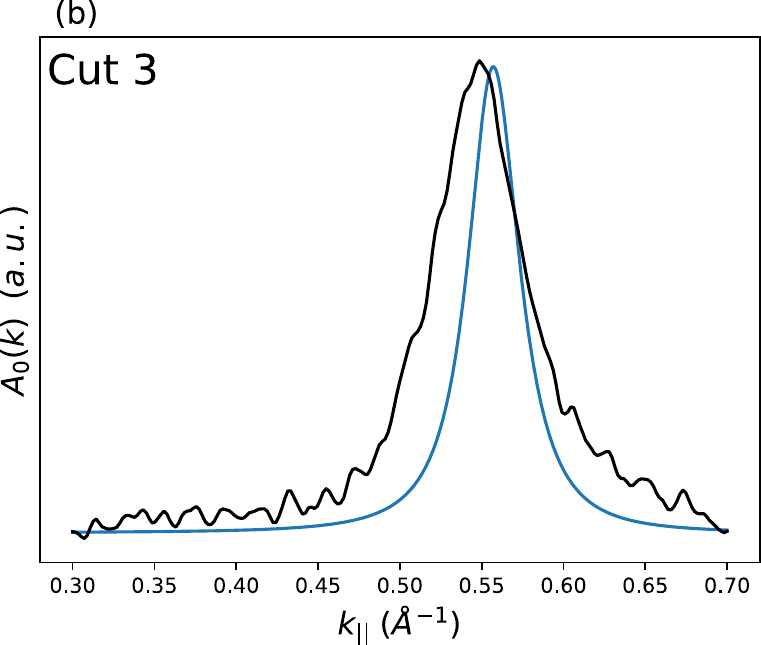}
    \phantomsubcaption\label{fig:4b}
\end{minipage}
\hfill
\begin{minipage}{.32\textwidth}
    \centering
    \includegraphics[width=\linewidth]{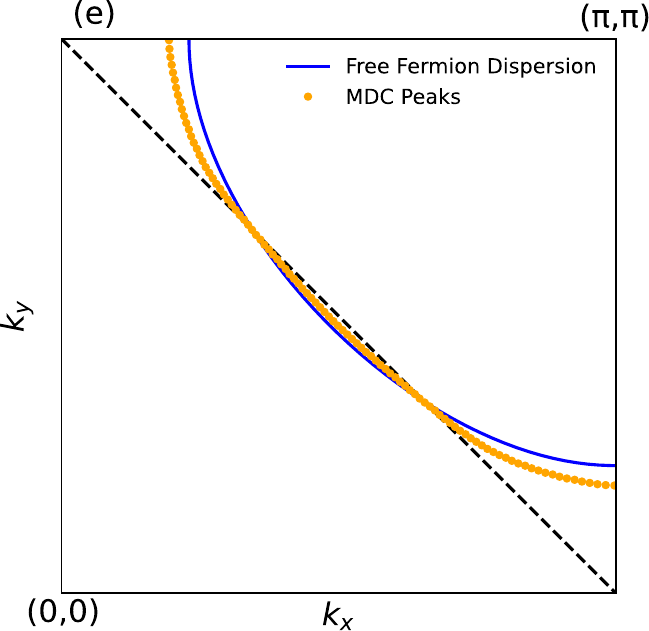}
    \phantomsubcaption\label{fig:4c}
\end{minipage}
 \vspace{0cm}
\begin{minipage}{.32\textwidth}
    \centering
    \includegraphics[width=\linewidth]{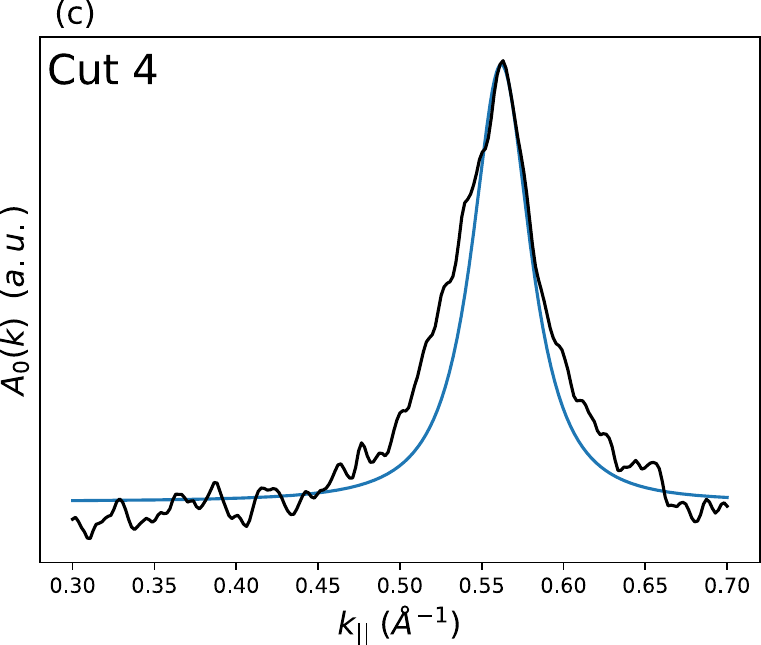}
    \phantomsubcaption\label{fig:4d}
\end{minipage}
\hfill
\begin{minipage}{.32\textwidth}
    \centering
    \includegraphics[width=\linewidth]{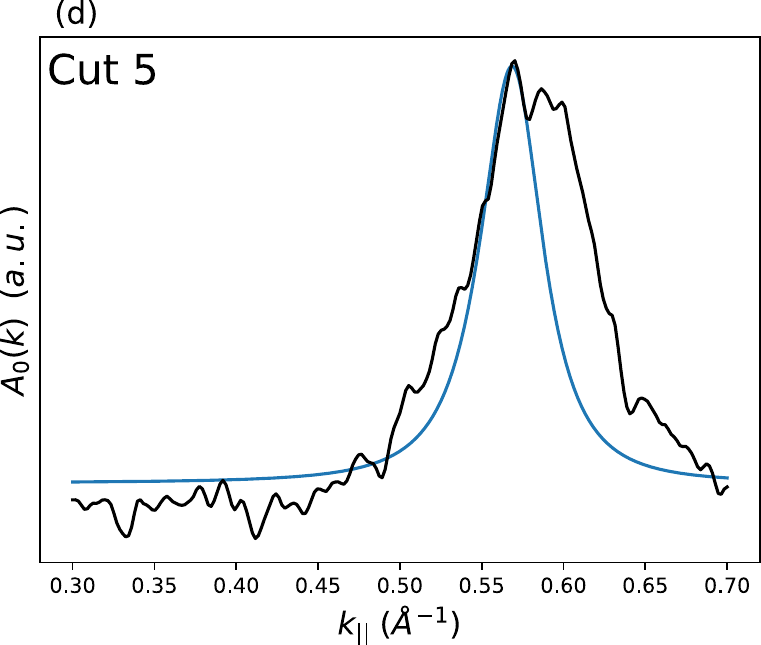}
    \phantomsubcaption\label{fig:4e}
\end{minipage}
\hfill
\begin{minipage}{.29\textwidth}
    \centering
    \includegraphics[width=\linewidth]{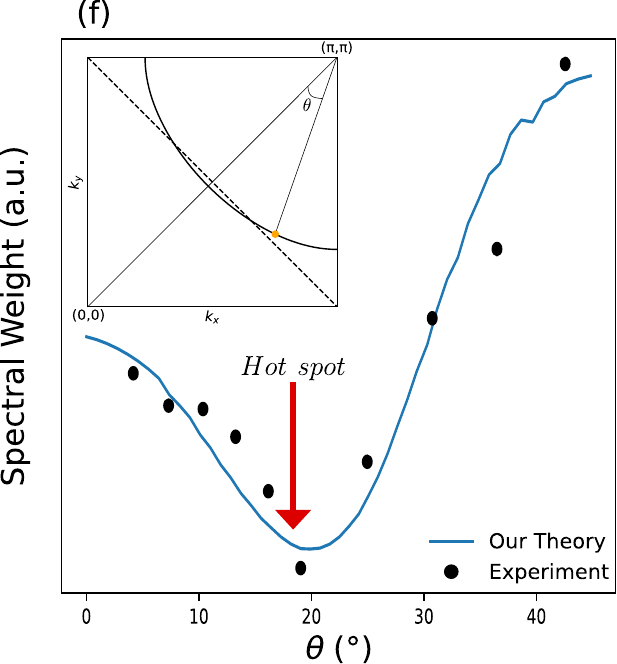}
    \phantomsubcaption\label{fig:4f}
\end{minipage}
\caption{(a)-(d) Experimental and theoretical MDC Spectra at zero frequency corresponding to cuts 2-5 respectively. (e) Free fermion dispersion vs actual positions of the MDC peaks at zero frequency along multiple diagonal cuts. (f) Experimental and theoretical integrated spectral weight in the range of $[-30,0] \; meV$ along the Fermi surface. The red arrow indicates the position of the hot spot. The discrepancy  between  our theory and the data is likely the result of our approximation of the fermionic self-energy by the perturbative one-loop formula.  At the two-loop level, the form of the theoretical spectral function is qualitatively similar to that at the one-loop order, but differs in detail (see Appendix \ref{AppendixB}).}
\label{fig:4}
\end{figure}
\begin{figure}[H]
\centering
\begin{minipage}{.32\textwidth}
    \centering
    \includegraphics[width=\linewidth]{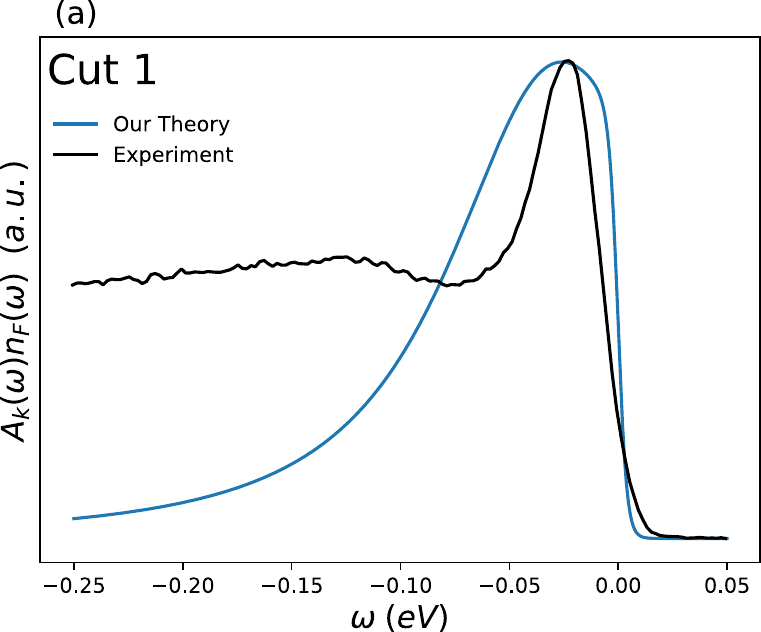}
    \phantomsubcaption\label{fig:5a}
\end{minipage}
\hfill
\begin{minipage}{.32\textwidth}
    \centering
    \includegraphics[width=\linewidth]{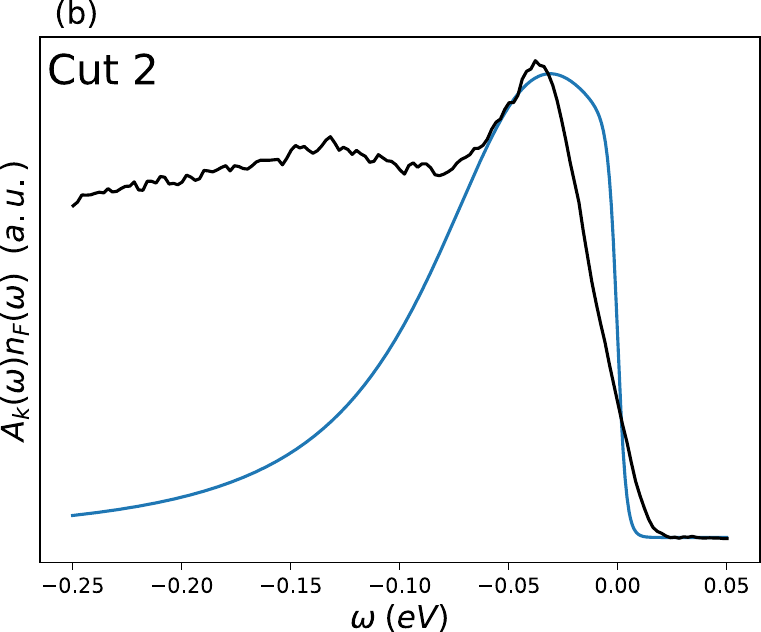}
    \phantomsubcaption\label{fig:5b}
\end{minipage}
\hfill
\begin{minipage}{.32\textwidth}
    \centering
    \includegraphics[width=\linewidth]{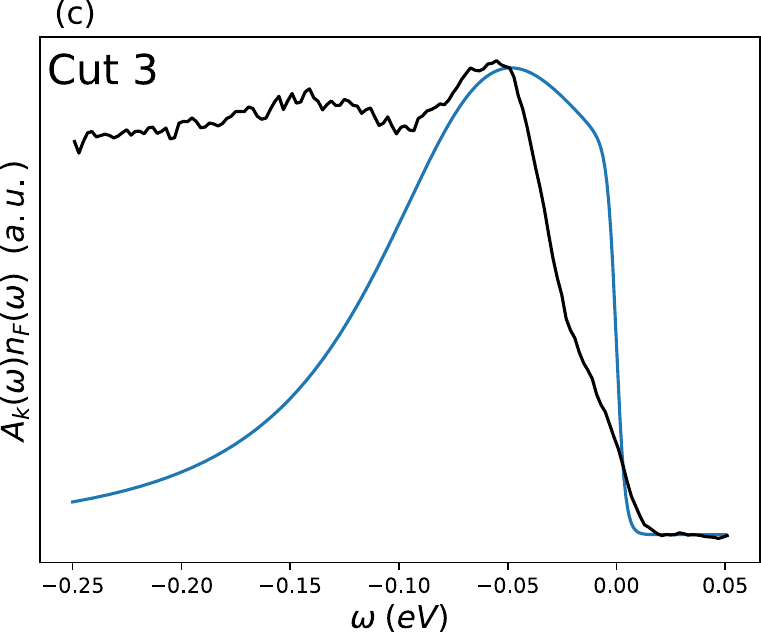}
    \phantomsubcaption\label{fig:5c}
\end{minipage}
 \vspace{0cm}
\begin{minipage}{.32\textwidth}
    \centering
    \includegraphics[width=\linewidth]{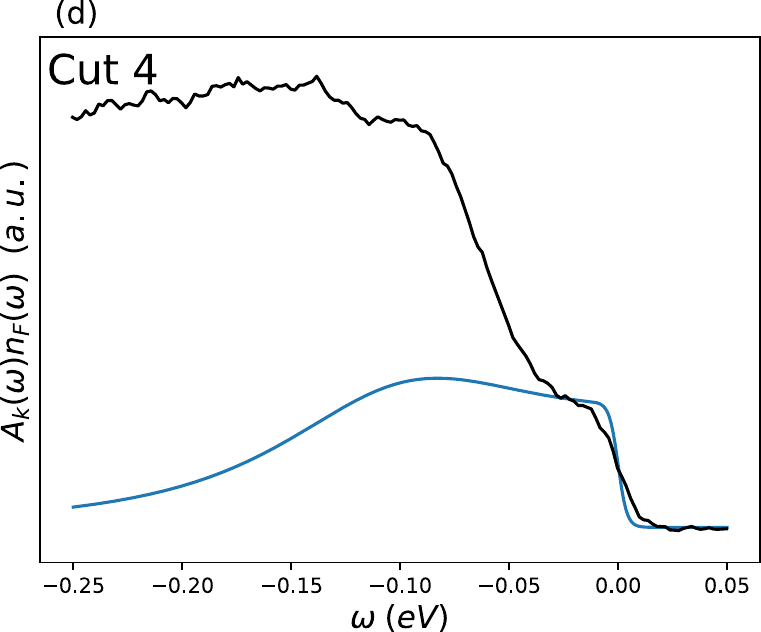}
    \phantomsubcaption\label{fig:5d}
\end{minipage}
\hfill
\begin{minipage}{.32\textwidth}
    \centering
    \includegraphics[width=\linewidth]{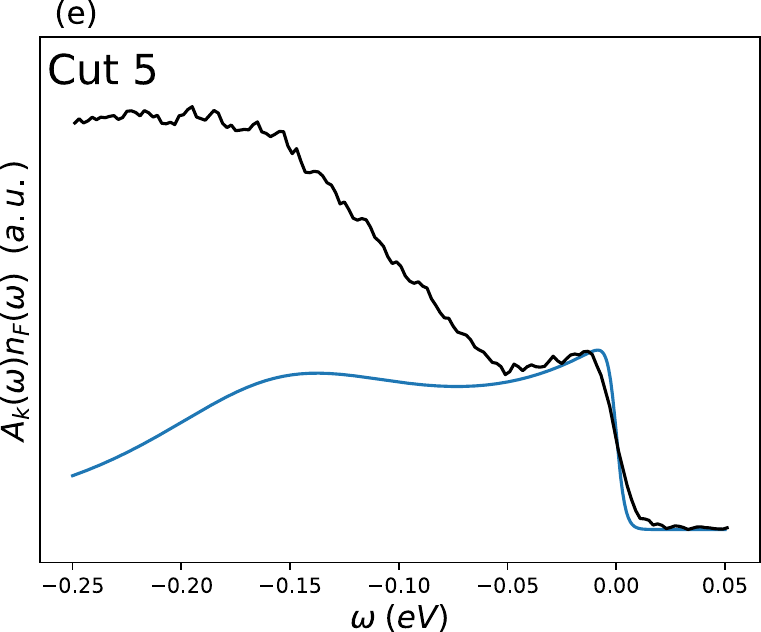}
    \phantomsubcaption\label{fig:5e}
\end{minipage}
\hfill
\begin{minipage}{.32\textwidth}
    \centering
    \includegraphics[width=\linewidth]{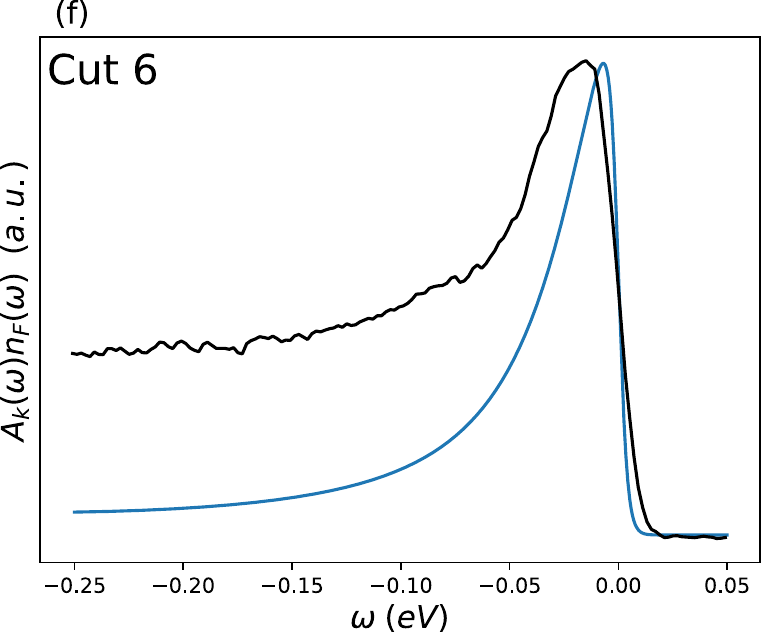}
    \phantomsubcaption\label{fig:5f}
\end{minipage}
\caption{(a)-(f) Experimental and theoretical EDC Spectra at Fermi momentum in cuts 1-6.  Observe that the theoretical EDC reproduces both the pseudogap peak and the peak in cut 5 near zero frequency. }
\label{fig:5}
\end{figure}

In Figs. \ref{fig:4} and \ref{fig:5},  we present some additional details of the comparison between our theoretical results and the experimental data reported in Ref.\cite{Xu_2023}. In  Figs. \ref{fig:4a}–\ref{fig:4d} we show MDC spectra at zero frequency along cuts 2–5. We  see that our theory  accurately captures both the peak positions and overall spectral shapes. A small discrepancy in cut 2 is likely due to deviations of the actual fermionic dispersion from $t-t'$ form as seen in \ref{fig:4e}. A similar deviation from the free fermion dispersion has previously been reported in \cite{Roy_2007}. In Fig.~\ref{fig:4f}, we  compare experimental and theoretical results for the spectral weight, defined in \cite{Xu_2023}  as the momentum-resolved integral of the spectral intensity over the energy window $[-30, 0]$~meV, along the underlying Fermi surface. Both theory and experiment show that the spectral weight reaches a minimum at the hot spot. The theoretical curve reproduces the experimental trend quite well, lending further support to the validity of our approach in capturing the essential features of the low-energy electronic structure. In Figs. \ref{fig:5a}–\ref{fig:5f} we compare our theory with the EDC data from \cite{Xu_2023} at $k=k_F$ for all six cuts. We compare peak positions and low-energy features, but do not aim to match the high-frequency behavior of $A_{k_F} (\omega)$ as the latter is heavily affected by the quantum self-energy $\Sigma_{qm} (\omega)$, not included into Eq. (\ref{2})
\footnote{We recall that for MDC spectrum, taken at a fixed $\omega$, $\Sigma_{qm}(\omega)$ does not modify the Lorentzian lineshape, but for EDC spectrum, the full frequency dependence of $\Sigma_{qm}(\omega)$ is essential, and it leads to a deviation of $A_{k_F} (\omega)$ from a Lorentzian lineshape and produces high-energy tails.}.
The latter broadens the EDC spectrum but {\it does not} give rise to pseudogap
\footnote{ The quantum self-energy  can be computed within Eliashberg theory as dynamical antiferromagnetic spin fluctuations are Landau overdamped and  for this reason are slow modes compared to fermions.}.
We see that the theoretical and experimental peak positions  match quite well for all cuts, although the  peaks are less pronounced for cuts 4 and 5.  We also note that in cut 4, the spectral intensity is flat at small negative $\omega$, and in cuts 5 and 6 it displays an additional peak near zero frequency; the peak is far stronger in cut 6. We discuss the origin of these low-frequency peaks in detail in Appendix \ref{AppendixE}. In short, the peaks emerge because for $k_F$  on the zone boundary side from $k_h$ (like in cuts 5 and 6), the spectral function is a non-symmetric function  of frequency: it has a stronger pseudogap peak at a positive $\omega$ and the peak position decreases as  $k_F$ approaches the zone boundary.  As a consequence, the spectral function at negative $\omega$ increases as $\omega$ approaches zero, with a slope which is larger in cut 6 than in cut 5. Combining the spectral function with the Fermi function $n_F (\omega)$, one immediately obtains the peak in the spectral intensity in cut 5  at small but finite negative $\omega$, and an even stronger peak in cut 6.

\section{Superconductivity} We next discuss superconductivity out of a pseudogap phase.  We recall that Xu et al found the largest value of a superconducting gap at a hot spot, where the spectral weight above $T_c$ is mostly depleted, and called this a paradox of spin-mediated pairing.  We argue that within our theory these results are actually the expected ones as thermal fluctuations almost cancel out in the equation for the pairing vertex (the gap equation for brevity), such that  the reduction  of the low-energy spectral weight induced by thermal spin fluctuations does not influence  the momentum dependence of a superconducting gap along the  Fermi surface.  Without thermal fluctuations, the pairing gap has a maximum at a hot spot~\cite{Abanov_2008,Hassan_2008,Wang_2013}.

A near-cancellation of the contributions from thermal fluctuations have been reported before~\cite{Millis_1988,Abanov_2008,Wang_2016,Wu_2019,*paper_2,Hauck_2020,Wang_2020}, for the gap equation obtained within the Eliashberg approximation, when one factorizes the momentum integration between fermions and the pairing boson (the exact cancellation holds when one additionally neglect the momentum dependence of the pairing vertex near a hot spot). The physical argument is that  for spin-singlet pairing, thermal fluctuations mimic non-magnetic impurities, and their near-cancellation is essentially the realization of the Anderson theorem~\cite{anderson}. In our case, it is crucial that thermal self-energy is obtained by performing the momentum integration exactly, without factorization. Still, we demonstrate that thermal self-energy nearly cancels out.

We measure the pairing vertex $\Phi (k, \omega)$ in units of $v_F \xi^{-1}$, like other variables and measure momentum in units of $\xi^{-1}$. We consider the momentum right at the hot spot, ${\bf k} = k_h$,  where the depletion of the spectral weight above $T_c$ is the largest.  To simplify the presentation, we first neglect quantum self-energy, as we did before, and consider the linearized gap equation for infinitesimally small $\Phi (k, \omega)$. We then show  that the near-cancellation of the thermal contribution holds  when we include $\Sigma_{qm}$ and analyze a non-linear gap equation.

The equation for $\Phi(k_h, \omega)$  can be most straightforwardly analyzed in Matsubara frequencies: fermionic  $\omega_m = \pi T (2m +1)$ and bosonic $\Omega_m = 2\pi T m$.  We search for d-wave superconductivity and set $\Phi_{k+Q, \omega_m} = - \Phi (k, \omega_m)$.  The gap equation is
   \beq
   \Phi(k_h, \omega_m) = \frac{3 {\bar g}}{(v_F \xi^{1})^2}
    T \sum_{\Omega_m} \int \frac{d^2q}{(2\pi)^2} \frac{\Phi (k_h+ q\xi^{-1}, \omega_m + \Omega_m)}{(\omega_m + \Omega_m + \Sigma_{th} (k+Q +q \xi^{-1},  \omega_m + \Omega_m))^2 + q^2_{\perp}} \chi (q+Q, \Omega_m),
  \label{4}
  \eeq
where $\chi (q+Q, \Omega_m)$ is the dynamical spin-fluctuation propagator, an extension of  $\chi (q+Q,0) =1/(q^2 + \xi^{-2})$. To single out the contribution from thermal spin fluctuations, we re-express $\Phi (k, \omega)$ as
   \beq
   \Phi (k, \omega_m) = \Delta (k, \omega_m) \frac{\omega_m + \Sigma_{th} (k, \omega_m)}{\omega_m}
   \label{5}
   \eeq
and single out the thermal piece with $\Omega_m =0$ in the summation over $\Omega_m$ in (\ref{4}). We then re-express the gap equation as
   \beq
   \Delta (k_h, \omega_m) (1 + J_{k_h})   = {\hat L}\left[ \Delta\right]
  \label{6}
  \eeq
where ${\hat L} \left[ \Delta\right] $ is the sum of the terms in the r.h.s. of (\ref{4}) with $\Omega_m \neq 0$, and $J_{k_h}$ is the contribution from thermal fluctuations.  Using the expression for $\Sigma_{th}$, Eq. (\ref{1}), we obtain to first order in $\lambda$ (the accuracy, with which we computed $\Sigma_{th}$ in (\ref{1})):
\beq
   J_{k_h}= \lambda_{th} \int \frac{d^2 q}{2\pi} \frac{1}{q^2_\parallel+ q^2_\perp + 1} \frac{1 -\frac{\Delta (k + q \xi^{-1}, \omega_m)}{ \Delta (k, \omega_m)}}{q^2_\perp + \omega^2_m}
\label{7}
\eeq
To estimate the magnitude of $J$, we borrow the results of the earlier analysis of the gap equation without the thermal piece: (i) that typical dimensional  $\omega_m$ in the gap equation without the contribution from thermal fluctuations are of the order of the  quantum coupling constant $\lambda_{qm}  \geq 1$, and (ii) that the gap is the largest at a hot spot and varies at a scale of order $k_h$. Using (i) to estimate typical momenta in (\ref{7}), we find  $q \xi^{-1}  \sim k_F \lambda_{qm}/(k_h \xi) \ll k_h$. For such momenta, $\Delta (k + q \xi^{-1}, \omega_m)/ \Delta (k, \omega_m) \approx 1$. Expanding this ratio in  $q \xi^{-1}$ and using (ii),  we find that  $J \sim 1/(k_h \xi)^2$ is a small number. The same result holds if we include quantum self-energy $\Sigma_{qm}$. The latter is well described by the Eliashberg theory and predominantly depends on $\omega$, $\Sigma_{qm} \approx \Sigma_{qm} (\omega)$. To single out thermal fluctuations, we follow \cite{Wu_2019,*paper_2} and express $\Phi (k, \omega_m) =\Delta^* (k, \omega_m) (\omega_m + \Sigma_{qm} (\omega_m) + \Sigma_{th} (k,\omega_m))/(\omega_m + \Sigma_{qm} (\omega_m))$ The thermal correction factor $J$ becomes
\beq
   J^*_{k_h}= \lambda_{th} \int\frac{d^2 q}{2\pi} \frac{1}{q^2_\parallel+ q^2_\perp + 1} \frac{1-\frac{\Delta^* (k + q \xi^{-1}, \omega_m)}{ \Delta^* (k, \omega_m)}}{q^2_\perp + (\omega^2_m + \Sigma_{qm} (\omega_m))^2 + (\Delta^* (k + q \xi^{-1}, \omega_m))^2}
\label{8}
\eeq
Estimating $J^*$ in the same way as before, we find that it is still small in $1/(k_h \xi)^2$, i.e., thermal fluctuation nearly cancel out from the gap equation. For $k$ close, but still different from $k_h$, the analysis of the thermal piece in the gap equation is more involved, but by continuity we fully expect that $J_k$ remains small for all momenta along the Fermi surface. The smallness of $J_{k_h}$ and $J^*_{k_h}$ implies that thermal fluctuations only weakly affect the gap equation at a hot spot despite that they give rise to a strong depletion of the spectral weight in the normal state. The dependence of the gap function along the Fermi surface then remains the same as without thermal fluctuations.
For the last case, the gap along the Fermi surface, $\Delta({\bf k}_F)$,  has a  maximum at a hot spot, at least  for weak and moderate quantum $\lambda_{qm} ({\bf k}_F)$~\cite{Abanov_2008}.

\section{Conclusions}  In this communication, we reported the results of our study of the  pseudogap behavior in a metal near an antiferromagnetic instability under the assumption that pseudogap behavior is caused by thermal magnetic fluctuations. We specifically analyzed weak pseudogap regime (pseudogap energy is smaller than $v_F \xi^{-1}$), where we  computed fermionic self-energy along the Fermi surface beyond Eliashberg approximation. We demonstrated that EDC and MDC spectral functions behave differently: EDC  displays a pseudogap behavior with peaks at a finite frequency at all momenta, which gives rise to a depletion of the density of states at the lowest energies,  the largest one at a hot spot. At the same time,  MDC peaks disperse within the pseudogap, and end up at zero frequency at a gossamer Fermi surface, which is close to the Fermi surface of free fermions. We analyzed magnetic-mediated superconductivity and showed that thermal fluctuations almost cancel out in the gap equation despite that thermal self-energy  is obtained beyond the Eliashberg approximation, without factorization of the momentum integration between fermions and spin fluctuations. As the consequence, the gap function retains the same form as if thermal fluctuations were absent and has a maximum at a hot spot.   Our results reproduce a number of features found the in the ARPES study of NCCO at $x =0.15$ (Ref. \cite{Xu_2023}). The reduction of the spectral weight at small energies should also affect transport properties of electron-doped cuprates.  Given a good agreement of our results with ARPES, we believe that the one-loop thermal self-energy can be also used for the computation of transport characteristics, e.g., optical resistivity.

An intriguing question is whether the same scenario applies to hole-doped cuprates. These systems are certainly more correlated than electron-doped cuprates, and the pseudogap energy at small hole doping  is larger that $v_F \xi^{-1}$, which places weakly doped hole cuprates into a strong pseudogap regime. The results of some numerical studies~\cite{Gunnarsson2015,*Gull2015,*Schafer2021,Krien_2021,Berg_2023,Simkovic_2024} have been interpreted as evidence for pseudogap induced by thermal spin fluctuations. The analytical study, however found~\cite{Ye_2023,*Ye_2023_1} that a strong pseudogap behavior it is not guaranteed to hold if the ground state is not antiferromagnetically ordered. An alternative scenario for a pseudogap, supported by magnetotransport measurements~\cite{Chan_2024}, is that it is  a phase with a topological order and  full reconstruction of the Fermi surface into pockets (see \cite{Bonetti_2025}  for recent detailed review of this and other scenarios for pseudogap in hole-doped cuprates). More studies are needed to settle on the origin of pseudogap in hole-doped cuprates.

{\it{\bf Acknowledgments.}} We thank  N. Bultinck, M. Christos, R. Fernandes, R. Greene, Q. Guo, P.A. Lee,  W. Metzner, S. Sachdev, J. Schmalian, M. Ye, K.-J. Xu and particularly  Z.-X. Shen and A-M Tremblay for fruitful discussions and feedback. The work was supported by  the National Science Foundation
grant NSF: DMR-2325357. EKK acknowledges additional  cost-of-living support by the Onassis Foundation Scholarship ID: F ZU 034-1/2024-2025.

\bibliography{Cuprates_Paper}
\newpage
\appendix

\section{A Self-Consistent One-Loop Theory}
\label{AppendixA}
In this Appendix we evaluate the self energy in the self-consistent one-loop theory (SCOLT). For thermal antiferromagnetic fluctuations, the  SCOLT was first implemented in \cite{Vilk1997}, where the authors numerically solved for the self energy for a given set of system parameters and found that no pseudogap behavior emerges. Y. Vilk subsequently obtained thermal self-energy analytically~\cite{Vilk_1997}.  We reproduce his analytical expression and use it to analyze the slope of the frequency dependence of the spectral function near $\omega= 0$. We verify that no pseudogap behavior emerges for any value of $\lambda_{th}$.

The thermal self-energy $\Sigma (k,\omega)$ in SCOLT is the convolution of the static antiferromagnetic  propagator $\chi (q+Q) \propto 1/(q^2 + \xi^{-2} (T))$ and  the full fermionic propagator $G(k+q +Q, \omega)$. We once again measure $\omega$, $\Sigma (k, \omega)$  and $G(k, \omega) = (\omega - \epsilon^{*}_k  - \Sigma (k, \omega))^{-1}$  in units of $v_F \xi^{-1}$ and $q$ in units of $\xi^{-1}$. For definiteness, we set the momentum $k$ to be at a hot spot $k_h$. The self-energy at this momentum
is given by
\bea
 \Sigma_{th} (k_h, \omega) &=& \frac{3 {\bar g} T}{(v_F \xi^{-1})^2} \int \frac{d^2 q}{(2\pi)^2} \frac{1}{\omega - \Sigma_{th} (k_h +q, \omega) - q_{\perp}} \frac{1}{q^2_{\perp} + q^2_{\parallel}+ 1}
 \label{yy_1}
 \eea

Eq. (\ref{yy_1}) is an integral equation in momentum.  To proceed with  the analytical treatment, we assume that momentum dependence of the self-energy near a hot spot is weak and neglect it, i.e., approximate $\Sigma_{th}(k,\omega)$ by $\Sigma_{th} (\omega)$.  This reduces (\ref{yy_1}) to a non-linear algebraic equation. Introducing $f(\omega) = \omega -\Sigma_{th} (\omega)$, we re-express (\ref{yy_1})  as

\begin{equation}
f(\omega)=\omega-\lambda_{th}\left[ \frac{\log{\left(f(\omega) + \sqrt{1+ f(\omega)^2}\right)}}{\sqrt{1 + f(\omega)^2}}-i \frac{\pi}{2 \sqrt{1 + f(\omega)^2}}\right].
\label{eq:A2}
\end{equation}
Expanding Eq. \ref{eq:A2} at small $\omega$, one finds $f(\omega)=a\omega+ib(1-c\omega^2)+...$, where $a$,$b$ and $c$ are positive coefficients satisfying the equations
\begin{equation}
\begin{cases}
\begin{array}{c}
b\sqrt{1-b^2}=\lambda_{th}\left[\frac{\pi}{2}+i\log\left(ib+\sqrt{1-b^2}\right)\right],\\
b\sqrt{b^2-1}=\lambda_{th}\log\left(b+\sqrt{b^2-1}\right),
\end{array} & \begin{array}{c}
\lambda_{th}<1\\
\lambda_{th}\geq1
\end{array}\end{cases}
\label{eq:A3}
\end{equation}
\begin{equation}
a=1+\frac{a}{1-b^2}(b^2-\lambda_{th})
\label{eq:A4}
\end{equation}
\begin{equation}
c=\frac{a^2}{2(1-b^2)}\frac{1+2b^2-3\lambda_{th}}{1-2b^2+\lambda_{th}}
\label{eq:A5}
\end{equation}

\begin{figure}[H]
\begin{center}
\includegraphics[width=.7\textwidth]{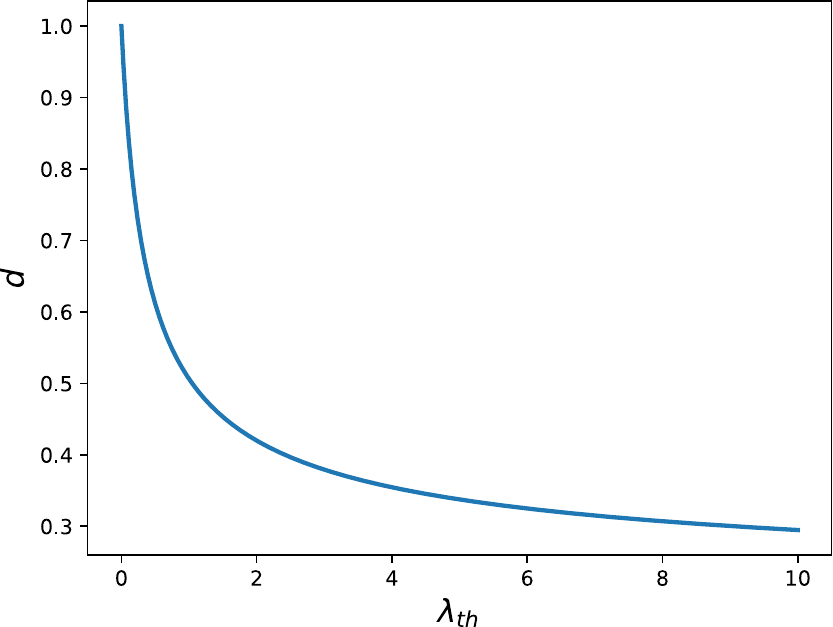}
\end{center}
\caption{Numerical evaluation of the quantity $d$ from Eq. \protect \ref{uu_1} as a function of the thermal coupling $\lambda_{th}$. This quantity remains positive for all values of $\lambda_{th}$, implying that no pseudogap develops.}
\label{fig:6}
\end{figure}

The spectral function at small frequencies can then be expressed as
\begin{equation}
A(k_h,\omega)=\frac{1}{\pi}|\text{Im}G(k_h,\omega)|\propto 1 -\frac{d}{b^2}\omega^2
\label{uu_1}
\end{equation}
where $d = a^2-b^2c$.
The condition for pseudogap to  develop is $d <0$. We  solve Eqs. \ref{eq:A3}-\ref{eq:A5} for various values of the thermal coupling and verify that $d$ stays positive for all values of $\lambda_{th}$. We  illustrate this in Fig. \ref{fig:6}. The implication is that within SCOLT, the spectral function retains the peak at $\omega=0$ no matter how large $\lambda_{th}$ is. We argue in the main text and in the Appendices \ref{AppendixB} and \ref{AppendixC} that this result is an artifact of neglecting vertex corrections. When vertex corrections are included at the same footing as the self-energy corrections, the spectral function does display a pseudogap behavior.

\newpage
\section{Two-loop corrections to the self-energy}
\label{AppendixB}

\begin{figure}[H]
\centering
\begin{minipage}{.45\textwidth}
    \centering
    \includegraphics[width=\linewidth]{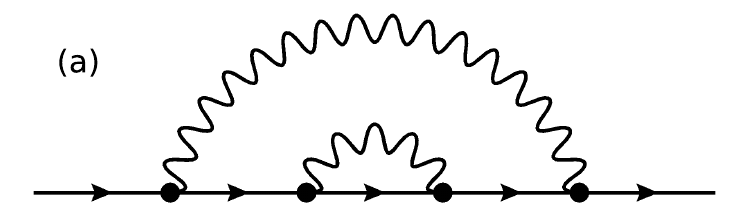}
    \phantomsubcaption\label{fig:7a}
\end{minipage}
\hfill
\begin{minipage}{.45\textwidth}
    \centering
    \includegraphics[width=\linewidth]{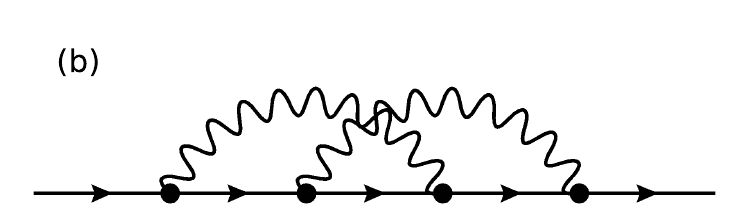}
    \phantomsubcaption\label{fig:7b}
\end{minipage}
\caption{(a) Non-crossed and (b) crossed two-loop diagrams for the self-energy.}
\label{fig:7}
\end{figure}

In this Appendix we analyze the two-loop thermal corrections to the fermionic self-energy. There are two diagrams at the two-loop order. One includes self-energy correction  to the propagator of an internal fermion (a non-crossed diagram), another includes a correction to an internal spin-fermion vertex (a crossed diagram).  The two diagrams are shown in Fig. \ref{fig:7}.

We again measure $\omega$, $\Sigma (k, \omega)$  in units of $v_F \xi^{-1}$ and $q$ in units of $\xi^{-1}$. For definiteness, we set the momentum $k$ to be at a hot spot $k_h$. The two terms in the two-loop self-energy at this momentum are given by

\begin{align}
\Sigma^{(2A)}_{th} (k_h,\omega)
&= \frac{\lambda_{th}^2}{4\pi^2}
   \int \frac{d^2q_1}{q_1^2+1}\,
        \frac{1}{(\omega - q_1\cos(\phi_1+\phi_0) + i\delta)^2} \int \frac{d^2q_2}{q_2^2+1}\,
        \frac{1}{\omega - q_1\cos\phi_1 - q_2\cos\phi_2 + i\delta}\nonumber\\
&= \frac{\lambda_{th}}{2\pi}
   \int \frac{d^2q_1}{q_1^2+1}\,
        \frac{1}{(\omega - q_1\cos(\phi_1+\phi_0) + i\delta)^2} \cdot\Sigma_{th}^{(1)}(k_h+q_1,\omega)
   \label{uu_4}
   \\[8pt]
\Sigma^{(2B)}_{th} (k_h,\omega)
&= -\frac{\lambda_{th}^2}{12\pi^2}
   \int \frac{d^2q_1}{q_1^2+1}\,
        \frac{1}{\omega - q_1\cos(\phi_1+\phi_0) + i\delta} \nonumber\\
&\quad\times
   \int \frac{d^2q_2}{q_2^2+1}\,
        \frac{1}{\omega - q_2\cos(\phi_2+\phi_0) + i\delta}\,
        \frac{1}{\omega - q_1\cos\phi_1 - q_2\cos\phi_2 + i\delta}
\label{uu_5}
\end{align}
where $\Sigma^{(2A)}_{th}(k_h,\omega)$ and $\Sigma^{(2B)}_{th}(k_h,\omega)$ are from the non-crossed and crossed diagrams, respectively, and $\Sigma^{(1)}_{th}(k,\omega)$  is the one-loop self-energy from the main text. The full two-loop self-energy  is $\Sigma^{(2)}_{th}(k_h,\omega) = \Sigma^{(2A)}_{th}(k_h,\omega)+\Sigma^{(2B)}_{th}(k_h,\omega)$. In deriving the expressions above, we used the approximations $\epsilon_{k_h+q}\approx v_F q\cos\phi$ and $\epsilon_{k_h+Q+q}\approx v_F q\cos(\phi+\phi_0)$, where $\phi_0=\pi/2+\arctan(v_{Fy}/v_{Fx})$. For our parameters, $\phi_0 =0.86 \pi$. The difference in the overall sign of the two diagrams arises from the summation over spin indices.

\begin{figure}[H]
\centering
\begin{minipage}{.45\textwidth}
    \centering
    \includegraphics[width=\linewidth]{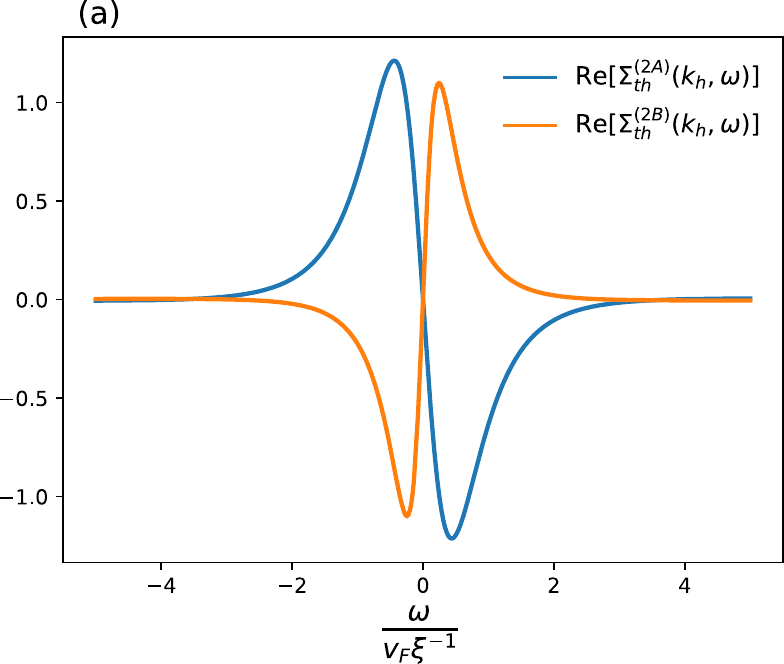}
    \phantomsubcaption\label{fig:8a}
\end{minipage}
\hfill
\begin{minipage}{.45\textwidth}
    \centering
    \includegraphics[width=\linewidth]{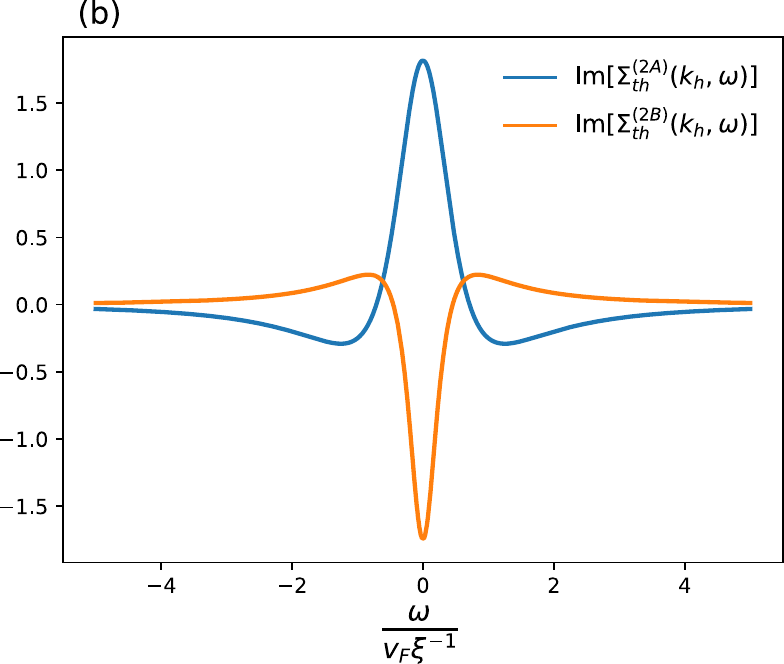}
    \phantomsubcaption\label{fig:8b}
\end{minipage}
\vspace{0cm}
\begin{minipage}{.45\textwidth}
    \centering
    \includegraphics[width=\linewidth]{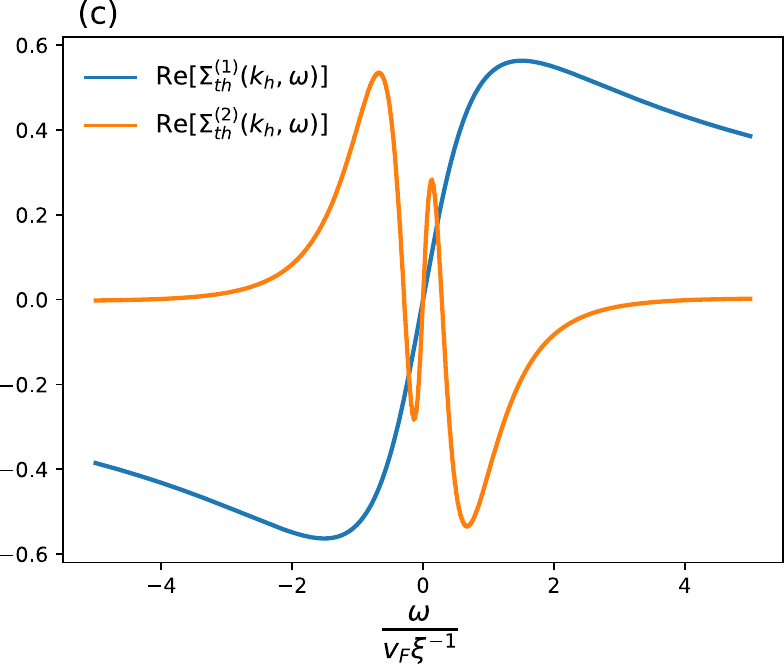}
    \phantomsubcaption\label{fig:8c}
\end{minipage}
\hfill
\begin{minipage}{.45\textwidth}
    \centering
    \includegraphics[width=\linewidth]{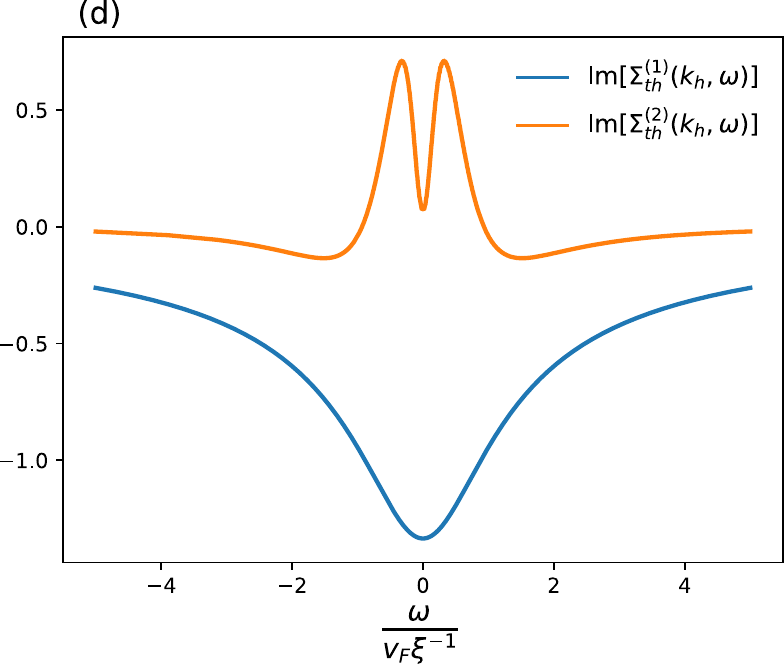}
    \phantomsubcaption\label{fig:8d}
\end{minipage}
\caption{(a) Real and (b) imaginary parts of the two-loop self-energy diagrams, with the contributions from each diagram shown separately. (c) Real and (d) imaginary parts of the one-loop and two-loop self-energies.}
\label{fig:8}
\end{figure}

\begin{figure}[H]
\centering
\begin{minipage}{.3\textwidth}
    \centering
    \includegraphics[width=\linewidth,height=4cm]{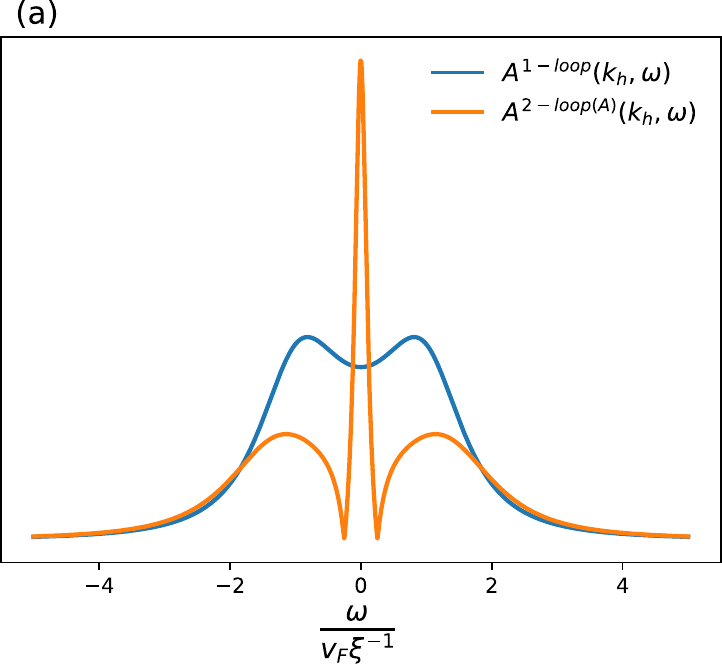}
    \phantomsubcaption\label{fig:9a}
\end{minipage}
\hfill
\begin{minipage}{.3\textwidth}
    \centering
    \includegraphics[width=\linewidth,height=4cm]{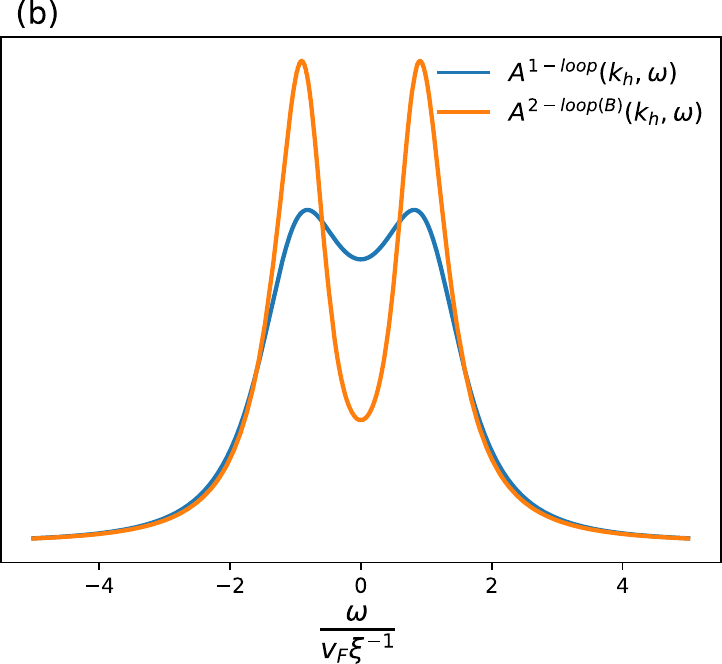}
    \phantomsubcaption\label{fig:9b}
\end{minipage}
\hfill
\begin{minipage}{.3\textwidth}
    \centering
    \includegraphics[width=\linewidth,height=4cm]{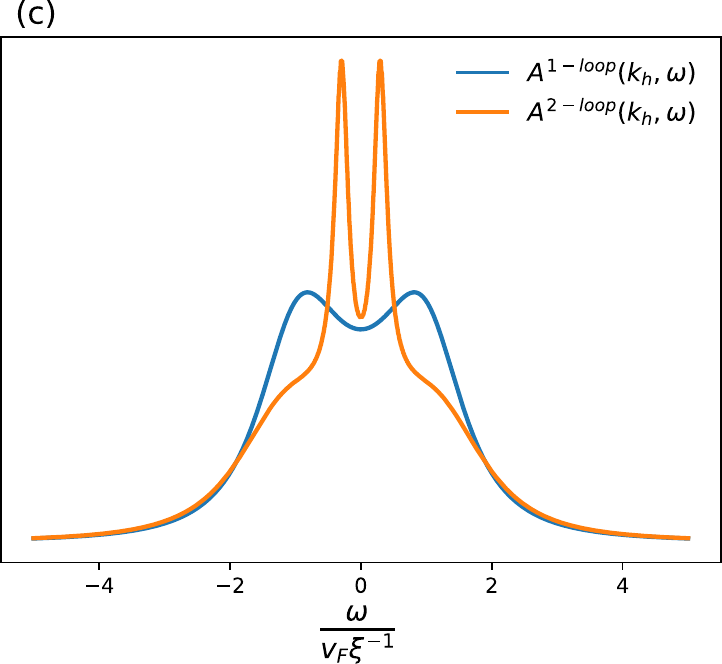}
    \phantomsubcaption\label{fig:9c}
\end{minipage}
\caption{(a) Spectral function including contributions from $\Sigma^{(1)}$ and $\Sigma^{(2A)}$. (b) Spectral function including contributions from $\Sigma^{(1)}$ and $\Sigma^{(2B)}$. (c) Spectral-function including all one-loop and two-loop self-energy corrections.}
\label{fig:9}
\end{figure}

We numerically evaluate the two-loop self-energy for $\lambda_{th}=0.85$ and show the results in Fig.~\ref{fig:8}. Taken separately, $\Sigma^{(2A)}_{th}$ and $\Sigma^{(2B)}_{th}$ actually exceed $\Sigma^{(1)}$ at small frequencies. However, the two contributions, taken separately, show different behavior. The self-energy $\Sigma^{(2A)}_{th}$  has the opposite sign at small frequencies compared to $\Sigma^{(1)}$. If we combine $\Sigma^{(1)}$ and $\Sigma^{(2A)}$, we find (see Fig. \ref{fig:9a}) the the spectral function develops a central peak at $\omega =0$.  This is an indication that higher-order rainbow diagrams for the self-energy (whose sum is a self-energy in a self-consistent one-loop approximation) tend to restore the spectral function with a single peak at $\omega =0$ (no pseudogap).  We discuss these series in some more details in the next Appendix. On the contrary, the two-loop self-energy $\Sigma^{(2B)}$ from the vertex correction diagram has the same sign as $\Sigma^{(1)}$ at small frequencies. Accordingly, if we combine $\Sigma^{(1)}$ and $\Sigma^{(2B)}$, we  find that the two-peak structure of the spectral function survives (see Fig. \ref{fig:9b}).
    
Comparing the magnitudes of $\Sigma^{(2A)}_{th}$ and $\Sigma^{(2B)}_{th}$, we find that they are essentially the same, despite that besides the sign change, the vertex correction diagram has an extra factor $(1/3)$ (see Eqs. (\ref{uu_4}) and (\ref{uu_5})).   As a result, the total self-energy $\Sigma^{(2)}$ is substantially smaller than each contribution. In  Fig. \ref{fig:9c} we plot the full spectral function at the two-loop order $A^{\text{2-loop}}(\omega)$.  We see that it preserves the two-peak structure, i.e., at least qualitatively, the two-loop result is similar to the one-loop one. The peaks in $A^{\text{2-loop}}(\omega)$  are at a lower frequency than in $A^{\text{1-loop}}(\omega)$.  This is consistent with the generic consideration in the next Appendix.  Higher-order diagrams will broaden the peaks in $A^{\text{2-loop}}(\omega)$  and make the actual spectral function more similar to the one-loop result. We therefore believe that  the one-loop description which we presented in the main text, captures the key effect of thermal spin fluctuations on the fermionic spectral function. 

\section{Role of vertex corrections for the pseudogap - general reasoning}
\label{AppendixC}

In this Appendix we present some general arguments why keeping vertex corrections to all orders in the expansion in $\lambda_{th}$ is essential for the description of the pseudogap. To simplify presentation, we consider a toy model with the static spin propagator approximated by a $\delta$- function at $\textbf{Q}$: $\chi(q) = \delta(\textbf{q}-\textbf{Q})$. For this $\chi (q)$, the one-loop thermal self-energy is
\begin{equation}
\Sigma^{(1)}(k_h,\omega)=3T {\bar g} \int \frac{d^2q}{(2\pi)^2}G^{(0)}(k+q,\omega)\chi(q)=\frac{3\lambda^2}{\omega+i\delta}
\end{equation}
where $\lambda\equiv\frac{\sqrt{\bar{g}T}}{2\pi}$ has units of energy. The one-loop spectral function is given by
\begin{equation}
A^{1-loop}(k_h,\omega)=
\frac{1}{\pi}\left|\text{Im}\left[\frac{1}{\omega+i\delta-\Sigma^{(1)}(k_h,\omega)}\right]\right|=
\frac{1}{2} \left(\delta (\omega - \sqrt{3} \lambda) + \delta (\omega + \sqrt{3} \lambda)\right)
\end{equation}

We see that the peak in a free-fermion Green's function at $\omega =0$ is split into two peaks at $\omega = \pm \sqrt{3} \lambda$ (Fig. \ref{fig:11a}). This is an extreme version of what we found for the one-loop spectral function in the main text, using an Ornstein-Zernike form of $\chi (q)$. 

We next compute the spectral function in the self-consistent one-loop theory, i.e., by including the infinite series of renormalizations of the propagator of an internal fermion, but neglecting vertex corrections. The full Green's function in this approximation is given by a quadratic algebraic equation
\begin{equation}
G^{-1}(k_h,\omega)=\omega+i\delta-\Sigma(k_h,\omega)=\omega+i\delta-3\lambda^2G(k_h,\omega)
\end{equation}
Solving it, we obtain at $\omega >0$, 
\begin{equation}
G(k_h,\omega)=\frac{\omega-\sqrt{(\omega+i\delta)^2-12\lambda^2}}{6\lambda^2}
\label{uu_7}
\end{equation}
The corresponding spectral function is
\begin{equation}
A(k_h,\omega)=\frac{1}{\pi}\left|\text{Im} G(k_h,\omega)\right|=\frac{\sqrt{12\lambda^2-\omega^2}}{6\pi\lambda^2}\Theta(12\lambda^2-\omega^2)
\end{equation}

\begin{figure}[H]
\begin{center}
\includegraphics[width=.7\textwidth]{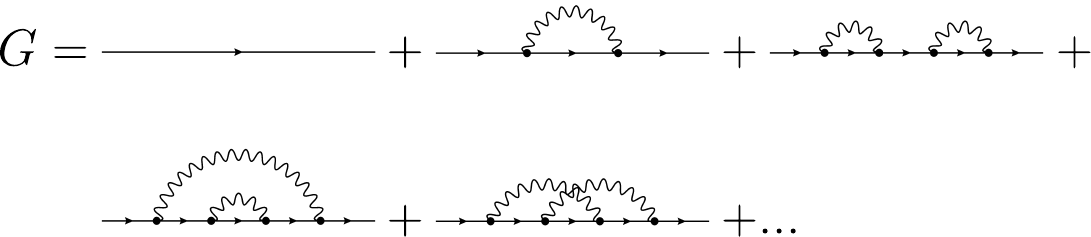}
\end{center}
\caption{Diagrammatic series for the full Green's function up to second order.}
\label{fig:10}
\end{figure}

\begin{figure}[H]
\centering
\begin{minipage}{.32\textwidth}
    \centering
    \includegraphics[width=\linewidth,height=4cm]{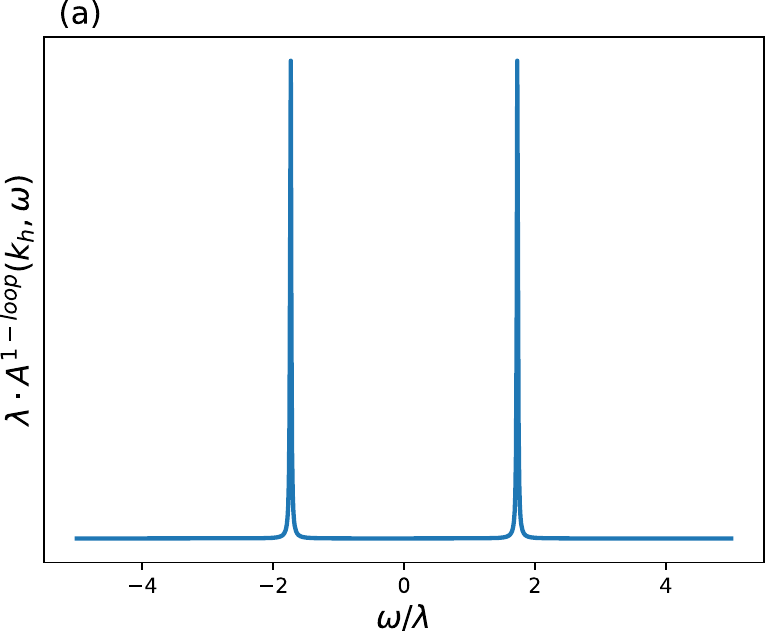}
    \phantomsubcaption\label{fig:11a}
\end{minipage}
\hfill
\begin{minipage}{.32\textwidth}
    \centering
    \includegraphics[width=\linewidth,height=4cm]{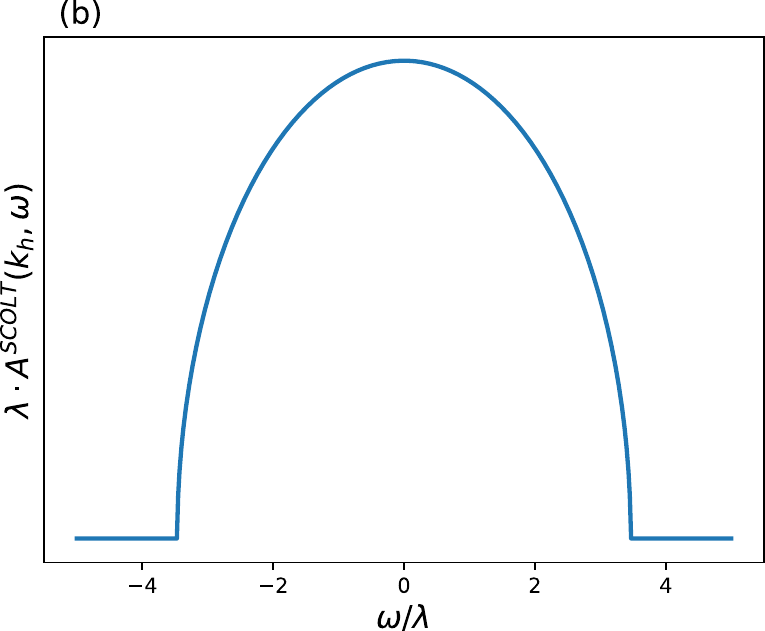}
    \phantomsubcaption\label{fig:11b}
\end{minipage}
\hfill
\begin{minipage}{.32\textwidth}
    \centering
    \includegraphics[width=\linewidth,height=4cm]{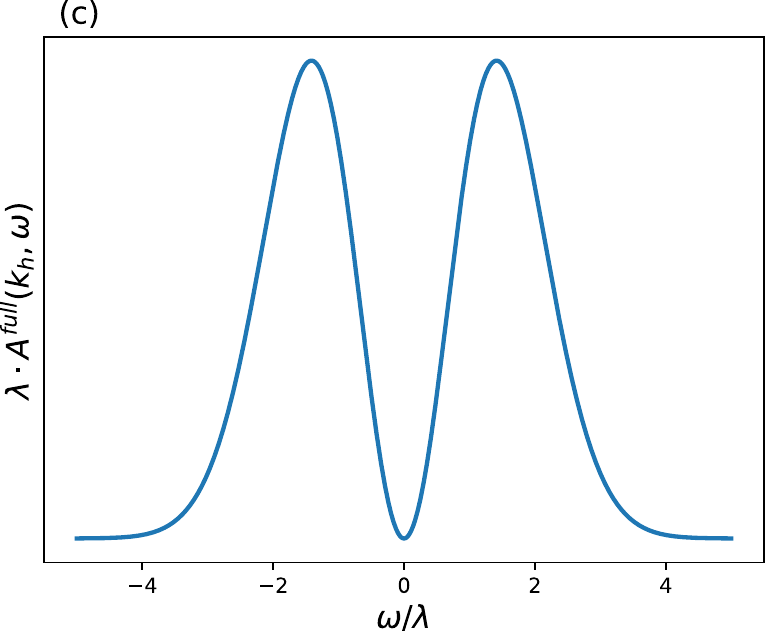}
    \phantomsubcaption\label{fig:11c}
\end{minipage}
\caption{(a) One-loop spectral function. (b) Spectral function from SCOLT. (c) Full spectral function.}
\label{fig:11}
\end{figure}

This spectral function has a single peak at $\omega =0$ (Fig. \ref{fig:11b}), i.e., it does display pseudogap behavior. We now  add vertex correction diagrams  and compute the full Green's function. The first few diagrams are presented in Fig. \ref{fig:10}. The series of this kind have been analyzed before \cite{Schmalian1999,Sadovskii_review,
Sedrakyan2010,Ye_2023,*Ye_2023_1} and we just cite the result
\begin{equation}
G^{full}(k_h,\omega)=\frac{1}{\omega+i\delta}+\frac{3\lambda^2}{(\omega+i\delta)^3}+\frac{15\lambda^4}{(\omega+i\delta)^5}+...=\sum_{n=0}^{\infty}\frac{(2n+1)!!\lambda^{2n}}{(\omega+i\delta)^{2n+1}}
\end{equation}
where the combinatoric factor $(2n+1)!!$ combines the number of diagrams and spin summation for each diagram. This series can be summed exactly using Borel summation procedure and the equality  $(2n+1)!!=\frac{2^{n+1}}{\sqrt{\pi}}\int_0^{\infty}e^{-t}t^{n+1/2}dt$.  A simple experimentation shows that the full $G (k_h, \omega)$ can be expressed as the integral

\begin{equation}
G^{full}(k_h,\omega)=\frac{2\omega}{\sqrt{\pi}}\int_0^{\infty}
dt
 \frac{\sqrt{t}e^{-t}}{(\omega+i\delta)^2-2\lambda^2t}
\end{equation}
Evaluating the integral, we obtain

\begin{equation}
G^{full} (k_h,\omega)= 
\frac{\sqrt{2} z}{\lambda} 
\left(\sqrt{\pi}ze^{-z^2}\text{Erfi}(z)-1-i\sqrt{\pi}ze^{-z^2}\right)
\end{equation}
where  $\text{Erfi}(x)=\frac{2}{\sqrt{\pi}}\int_0^xe^{t^2}dt$ is the imaginary error function  and $z\equiv\omega/(\sqrt{2}\lambda)$. The full spectral function is given by
\begin{equation}
A^{full}(k_h,\omega)=\frac{1}{\pi}\left|\text{Im} G^{full}(k_h,\omega)\right|=\sqrt{\frac{\pi}{2}}\frac{\omega^2}{\lambda^3}e^{-\frac{\omega^2}{2\lambda^2}}
\end{equation}

This spectral function again displays two peaks at frequencies $\omega=\pm\sqrt{2}\lambda$ (Fig. \ref{fig:11c}). The peak positions are at somewhat smaller frequencies than in the perturbative one-loop approximation, but overall the structure of $A^{full}$ is qualitatively similar to one-loop result (the key difference is that the full spectral function is  non-zero at all frequencies, so the functional form in Fig. \ref{fig:11c} is literally a pseudogap behavior).
  
\newpage
The outcome of this analysis is that  vertex corrections are crucial for the description of the pseudogap. In the extreme case that we considered,  vertex corrections do not cancel out higher-order self-energy corrections (the two-loop vertex correction diagram is $-1/3$ of the two-loop rainbow diagram), yet they reduce the magnitude of the self-energy at each higher-loop order, and as the exact solution shows, this is enough to preserve the  pseudogap behavior that shows up  at the one loop order.

For comparison with the two-loop analysis in the previous Appendix, we take one more look at the self-energy in the self-consistent one-loop approximation, Eq. (\ref{uu_7}), but keep fermionic damping finite, i.e., replace $\omega$ by $\omega + i \gamma$. This was argued~\cite{Millis1998,imp_cuprates2} to be a proxy to the case when the magnetic correlation length is finite. Including $\gamma$ into (\ref{uu_7}) and expanding to order $\lambda^4$, we obtain the self-energy
\beq
\Sigma (\omega) = \frac{3\lambda^2}{\omega + i \gamma} + \frac{9\lambda^4}{(\omega + i \gamma)^3} 
\label{uu_9}
\eeq

\begin{figure}[H]
\centering
\begin{minipage}{.32\textwidth}
    \centering
    \includegraphics[width=\linewidth,height=4cm]{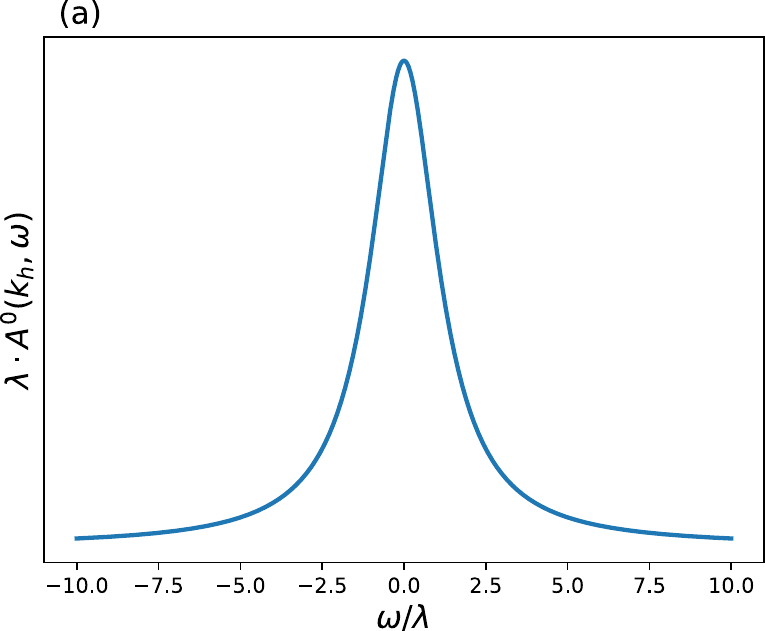}
    \phantomsubcaption\label{fig:12a}
\end{minipage}
\hfill
\begin{minipage}{.32\textwidth}
    \centering
    \includegraphics[width=\linewidth,height=4cm]{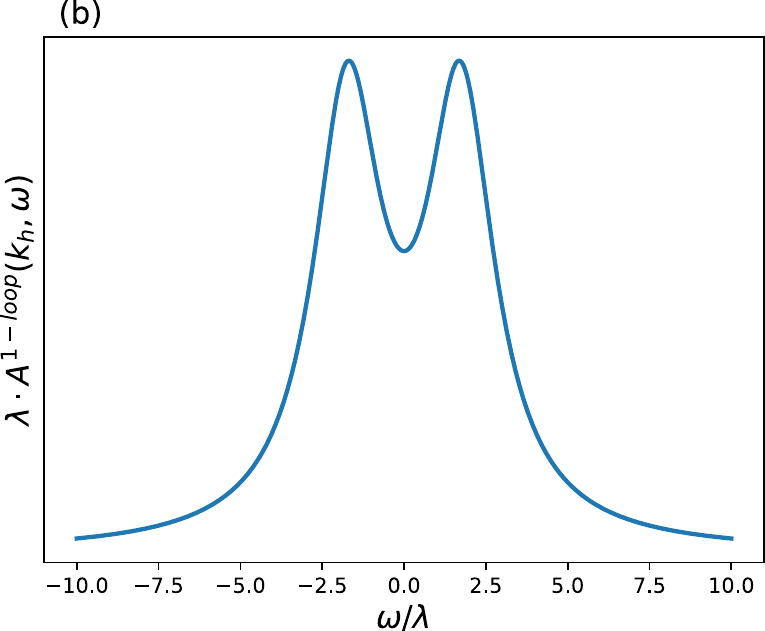}
    \phantomsubcaption\label{fig:12b}
\end{minipage}
\hfill
\begin{minipage}{.32\textwidth}
    \centering
    \includegraphics[width=\linewidth,height=4cm]{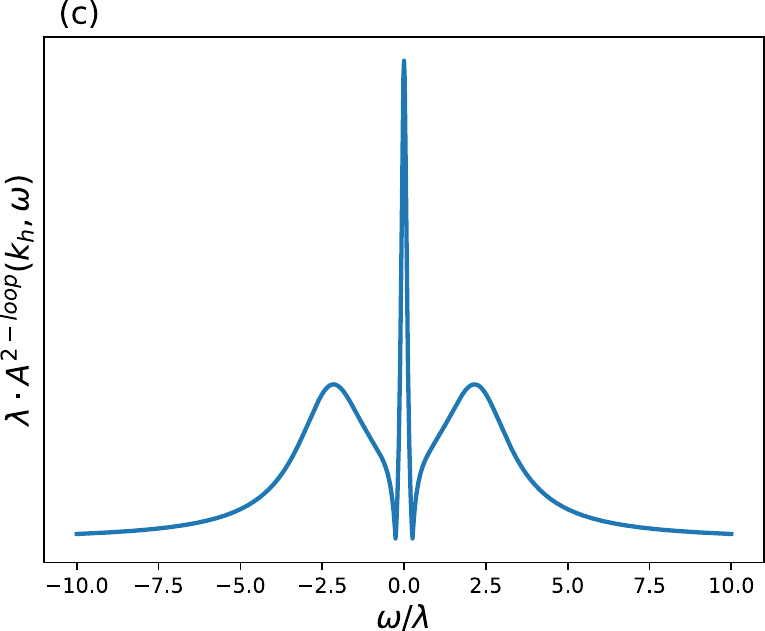}
    \phantomsubcaption\label{fig:12c}
\end{minipage}
\caption{(a) Zeroth order spectral function. (b) One-loop spectral function. (c) Two-loop spectral function only including the contribution from the non-crossed diagram. For all three plots we used $\gamma/\lambda=1.25$ for the broadening.}
\label{fig:12}
\end{figure}

At small frequencies $|\omega| < \gamma$, the two terms in (\ref{uu_9}) have opposite signs of both real and imaginary parts. This matches the discussion of our self-energy in the previous Appendix. We plot the spectral function in Fig. \ref{fig:12}. The bare spectral function $A^{0} (\omega)$ has a peak at $\omega =0$ with width $\gamma$ and the spectral function with one-loop self-energy $A^{\text{1-loop}} (\omega)$ has two peaks, if $\lambda > \gamma/\sqrt{3}$, like the one-loop spectral function in our case.  The two-loop spectral function  $A^{\text{1-loop}} (\omega)$, however, develops a peak at $\omega =0$. This is fully consistent with the results for our self-energy if we only include the rainbow diagram at the two-loop order.

\newpage
\section{Comparison of Theory and Experiment for cut 6}
\label{AppendixD}

\begin{figure}[H]
\centering
\begin{minipage}{0.66\textwidth}
    \centering
    \begin{minipage}{0.48\textwidth}
        \includegraphics[width=\linewidth]{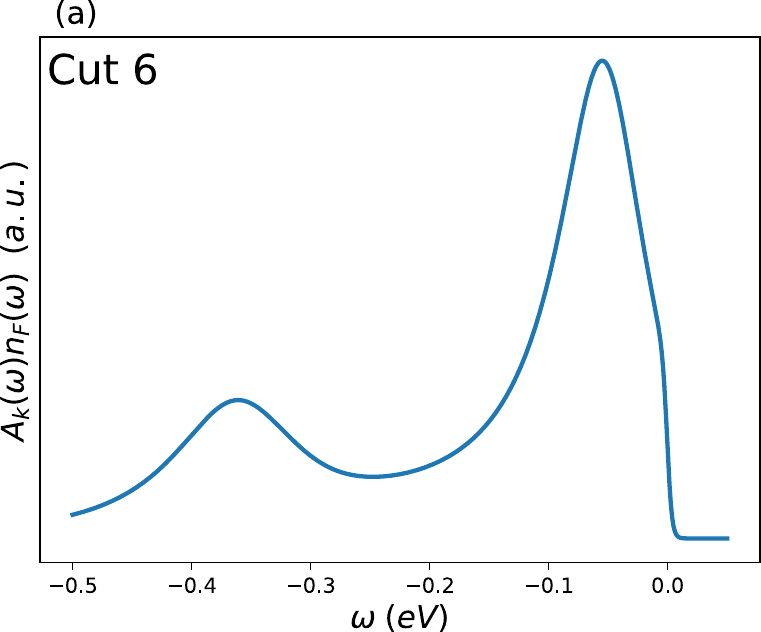}
        \phantomsubcaption\label{fig:13a}
    \end{minipage}
    \hfill
    \begin{minipage}{0.48\textwidth}
        \includegraphics[width=\linewidth]{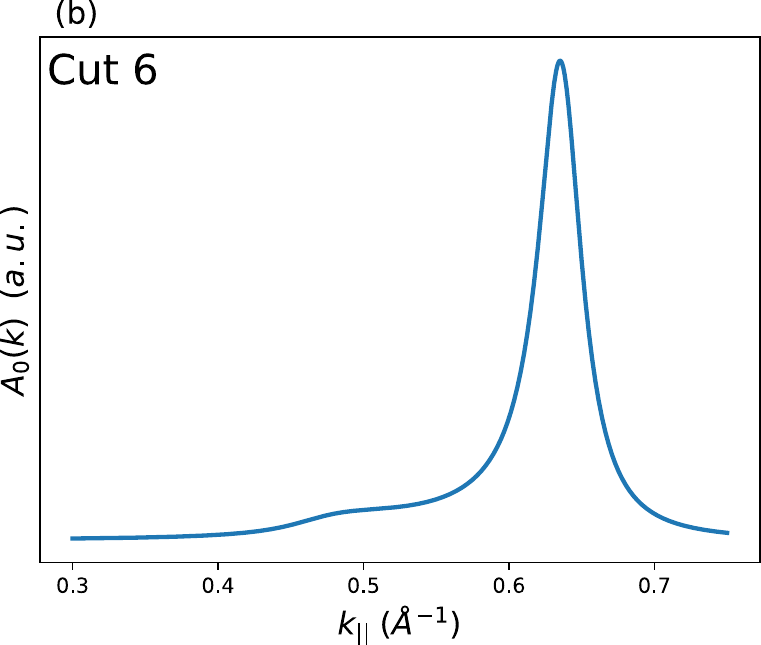}
        \phantomsubcaption\label{fig:13b}
    \end{minipage}
\end{minipage}

\vspace{0cm}
\begin{minipage}{.32\textwidth}
    \centering
    \includegraphics[width=\linewidth]{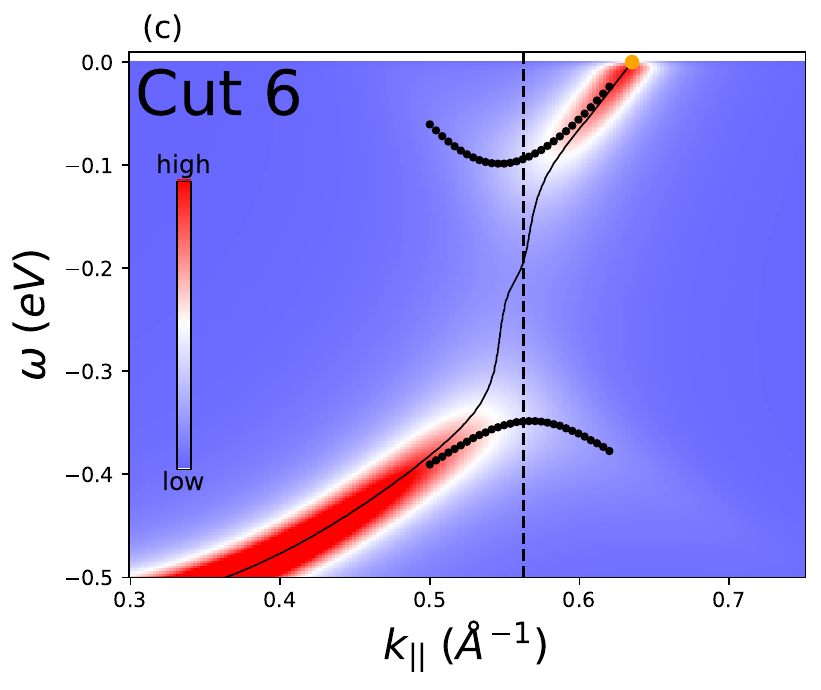}
    \phantomsubcaption\label{fig:13c}
\end{minipage}
\hfill
\begin{minipage}{.32\textwidth}
    \centering
    \includegraphics[width=\linewidth]{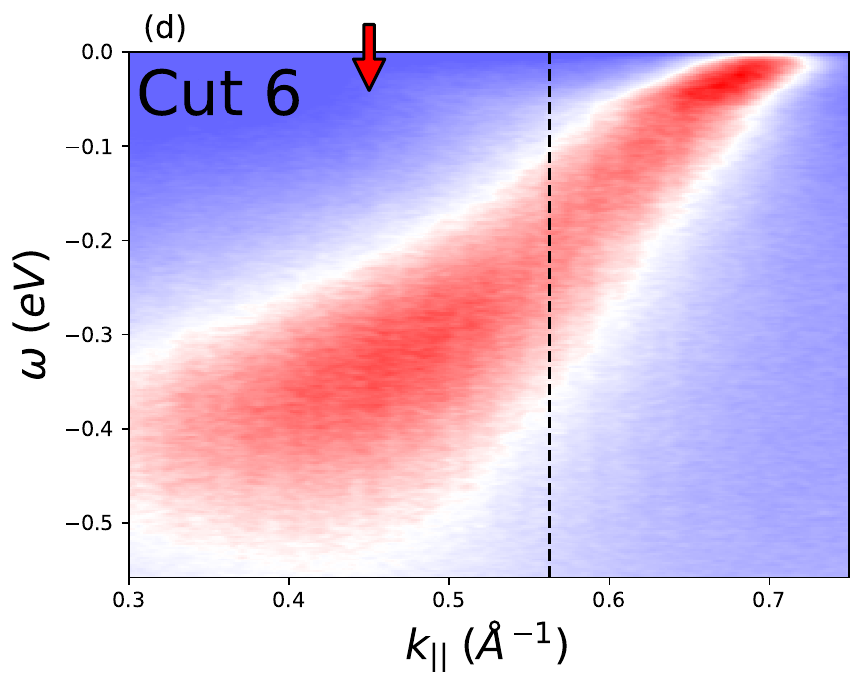}
    \phantomsubcaption\label{fig:13d}
\end{minipage}
\hfill
\begin{minipage}{.32\textwidth}
    \centering
    \includegraphics[width=\linewidth]{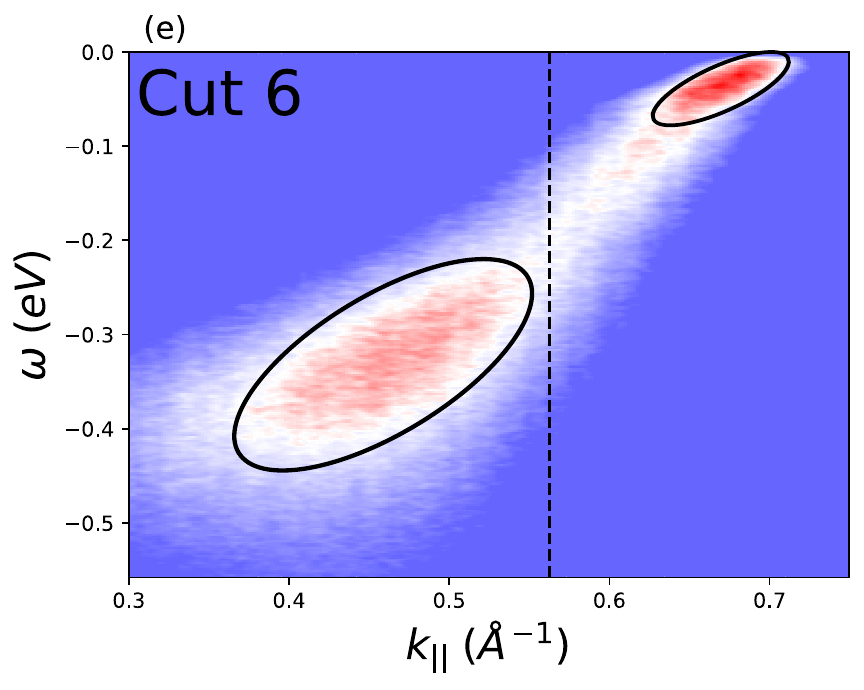}
    \phantomsubcaption\label{fig:13e}
\end{minipage}
\caption{(a) EDC spectrum of cut 6 at $k_{||}=0.6 \AA^{-1}$. (b) MDC spectrum of cut 6 for $\omega=0$. (c) Theoretical Frequency-Momentum Spectrum along cut 6. The black dots correspond to the positions of the EDC peaks for different values of momentum. The black line tracks the position of the MDC peak at different frequencies, with the orange dot representing $k_F$. The dashed line represents the magnetic Brillouin zone boundary. (d) Experimental Frequency-Momentum Spectrum for cut 6. The red arrow highlights the weak $(\pi,\pi)$-folded dispersion branch near the AF zone boundary that Xu et al. observe. (e) Experimental spectrum presented with a narrower color scale to enhance contrast. High-intensity regions are enclosed by ellipses for visual emphasis.}
\label{fig:13}
\end{figure}

In Fig. \ref{fig:13} we compare our theoretical results with experimental data along the momentum cut 6 (see Fig. \ref{fig:1}). This cut is for momenta near $(0,\pi)$. In the $(\pi,\pi)$ ordered state, this cut intersects the electron pocket centered at $(0,\pi)$.  We  compare  the theory and the data  over a broad energy range.

\begin{figure}[H]
\centering
\begin{minipage}{.32\textwidth}
    \centering
    \includegraphics[width=\linewidth]{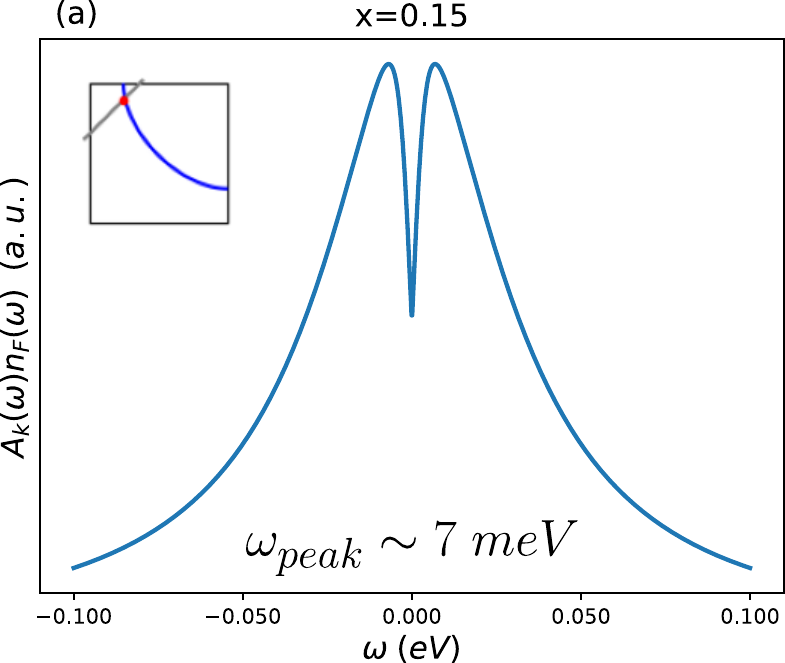}
    \phantomsubcaption\label{fig:14a}
\end{minipage}
\hfill
\begin{minipage}{.32\textwidth}
    \centering
    \includegraphics[width=\linewidth]{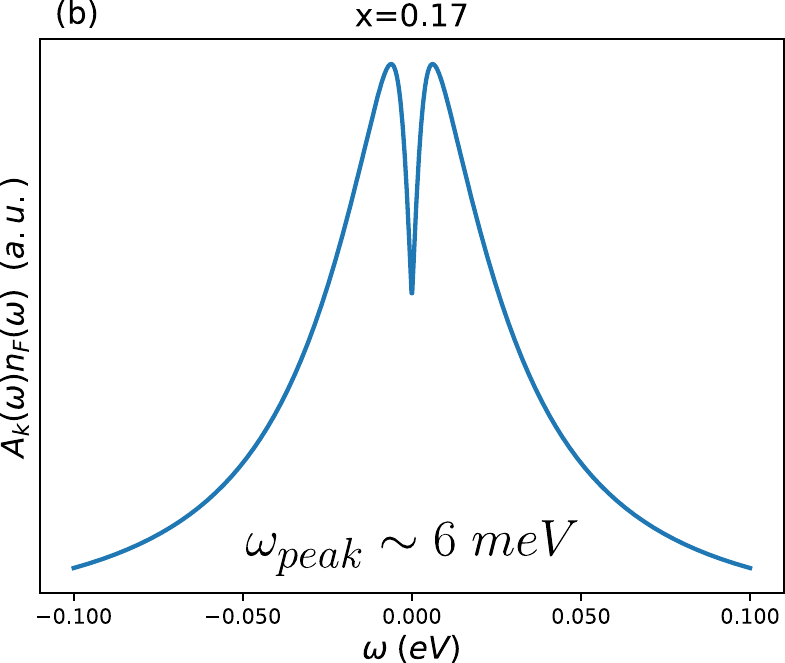}
    \phantomsubcaption\label{fig:14b}
\end{minipage}
\hfill
\begin{minipage}{.32\textwidth}
    \centering
    \includegraphics[width=\linewidth]{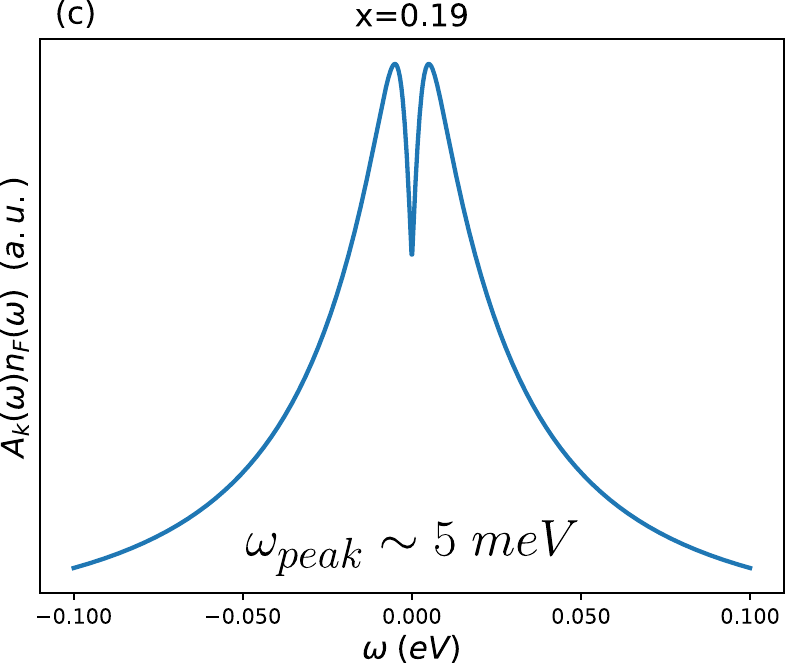}
    \phantomsubcaption\label{fig:14c}
\end{minipage}
 \vspace{0cm}
\begin{minipage}{.32\textwidth}
    \centering
    \includegraphics[width=\linewidth]{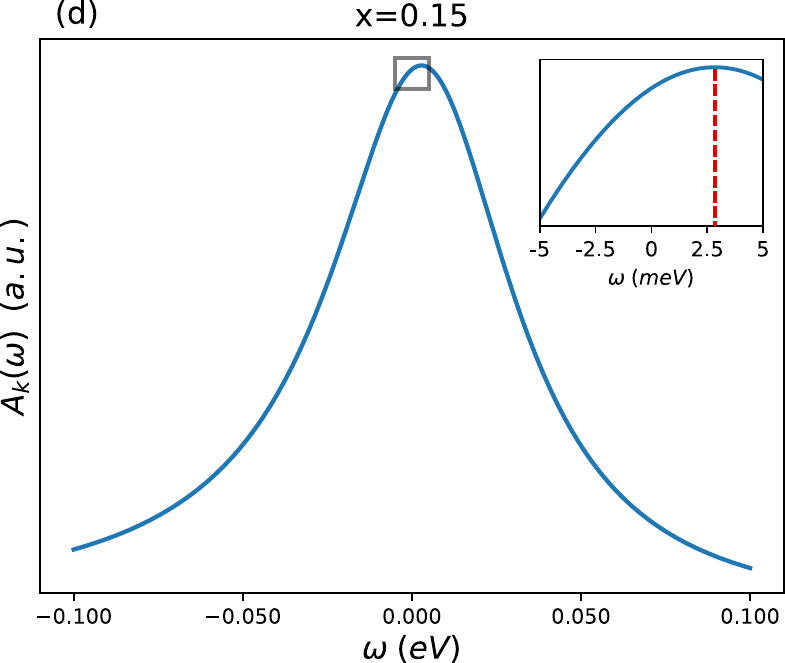}
    \phantomsubcaption\label{fig:14d}
\end{minipage}
\hfill
\begin{minipage}{.32\textwidth}
    \centering
    \includegraphics[width=\linewidth]{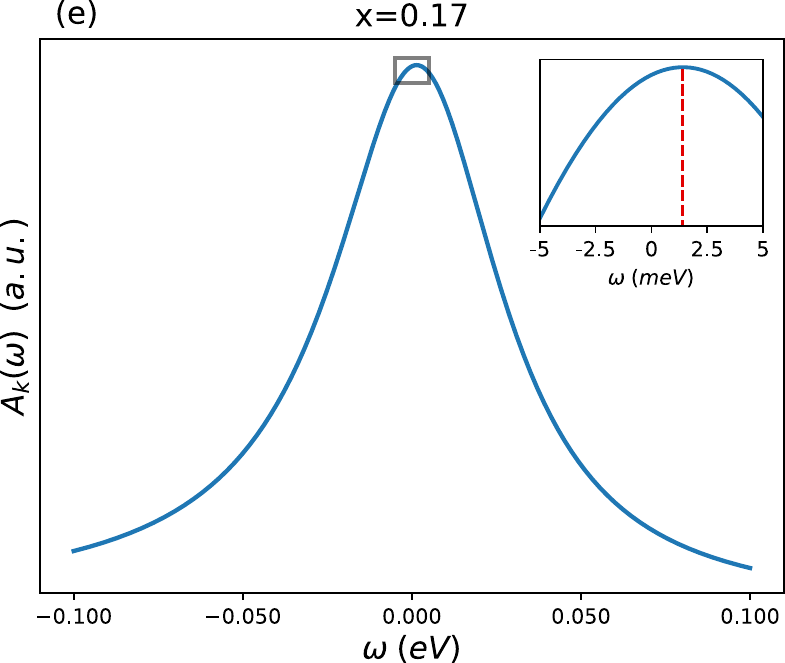}
    \phantomsubcaption\label{fig:14e}
\end{minipage}
\hfill
\begin{minipage}{.32\textwidth}
    \centering
    \includegraphics[width=\linewidth]{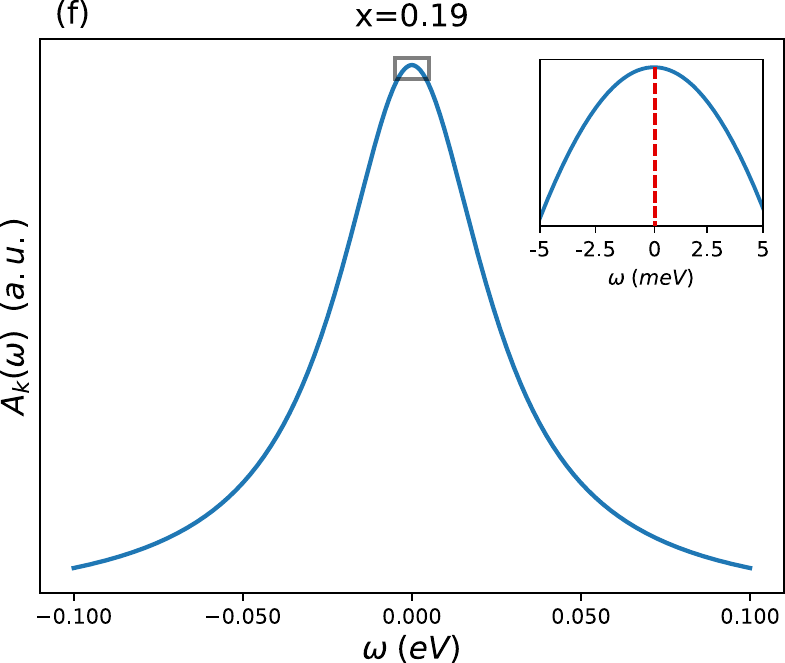}
    \phantomsubcaption\label{fig:14f}
\end{minipage}
\caption{(a)-(c) Symmetrized EDC spectral intensities at Fermi momentum (corresponding to the cut shown in the inset) for different values of the electron doping. We set $T=25K$ and symmetrized the intensity, originally obrtained at negative $\omega$,  for better comparison with Ref.\cite{Xu_2024}, where the symmetrized data have been presented.   (d)-(f) EDC spectral function at Fermi momentum for the same values of the electron doping. The insets show the behavior at small $\omega$. In the pseudogap phase, there is a peak at a small positive frequency and another peak at a large negative frequency (not shown). }
\label{fig:14}
\end{figure}

We show the theoretical spectrum  in Fig. \ref{fig:13c}. As we see from Fig. \ref{fig:13a}, it exhibits two EDC pseudogap peaks at negative frequencies, both of which should, in principle, be observable in experiment. These two peaks are descendants of two bands in the $(\pi,\pi)$ ordered state -- a hole-like band at large negative energies and an electron-like band that crosses the Fermi level. In Fig. \ref{fig:13d} we plot the experimental data from \cite{Xu_2023}. The data show weak $(\pi,\pi)$-folded dispersion branch near the antiferromagnetic zone boundary,  highlighted by the red arrow. It is consistent with the back-bending behavior of the lower-frequency EDC peak in the theoretical spectrum. To better resolve subtle features in the experimental data, we re-plot the experimental spectrum in Fig. \ref{fig:13e} using a more restricted color scale. This narrower color mapping range enhances the visibility of spectral intensity variations and reveals that the spectral weight is concentrated in two distinct regions, in agreement with the theoretical Fig. \ref{fig:13c}. Between these regions, the spectral intensity is notably suppressed.

\subsection{Doping evolution of the low-energy peak}

In Ref. \cite{Xu_2024}, Xu et al presented the results for the doping evolution of the doping  evolution of the lower-energy EDC peak in cut 6  at a fixed temperature. In our theory, the doping-dependent parameters are the chemical potential and the correlation length $\xi (x)$. To compare the theory and the experiment, we set $T = 25K$, as in Ref. \cite{Xu_2024}, approximate the chemical potential by its value for free fermions $\mu = \mu_0(x)$, and choose the correlation lengths $\xi_{x=0.15} = 10a$, $\xi_{x=0.17} = 5a$, and $\xi_{x=0.19} = 2a$, to model a progressive reduction of magnetic correlations with increasing doping. For these parameters, the doping levels $x = 0.15$ and $x = 0.17$ fall within the pseudogap regime, while for  $x = 0.19$ the system is in a conventional metallic state.

We show theoretical results for the low-frequency peak in Fig. \ref{fig:14}. We plot symmetrized spectral intensity to compare with the data in Fig. 3B of Ref. \cite{Xu_2024}. We see that the frequency of the low-energy peak in the spectral intensity monotonically decreases as the systems moves towards the metallic state. We note that it remains finite even at larger dopings, when the spectral function displays a conventional metallic behavior with the peak at $k_F$ located at zero frequency. This is just the effect of the Fermi function, which cuts the spectral intensity at the smallest negative $\omega$.
We discuss a similar effect in the next Appendix. The analysis of doping values $x<0.15$ is not possible within our one-loop perturbation theory because for $x<0.15$ the thermal coupling becomes larger than one implying that the system crosses over to a strong pseudogap regime.

\newpage
\renewcommand{\thesubsection}{\Alph{section}\arabic{subsection}}
\section{The low-energy peak in the EDC spectrum  in cut 5}
\label{AppendixE}

We see from Fig. \ref{fig:5e} that the  experimental EDC spectra  at $k_F$ for cut 5 possess a maximum  at a low energy in addition to a maximum at a larger negative $\omega$, associated with the pseudogap. The low-energy peak was further analyzed in ~\cite{Xu_2024}) and attributed to superconducting fluctuations.  We argue that such a peak naturally arises in our theory of pseudogap due to thermal spin fluctuations.

\begin{figure}[H]
\centering
\begin{minipage}{.45\textwidth}
    \centering
    \includegraphics[width=\linewidth]{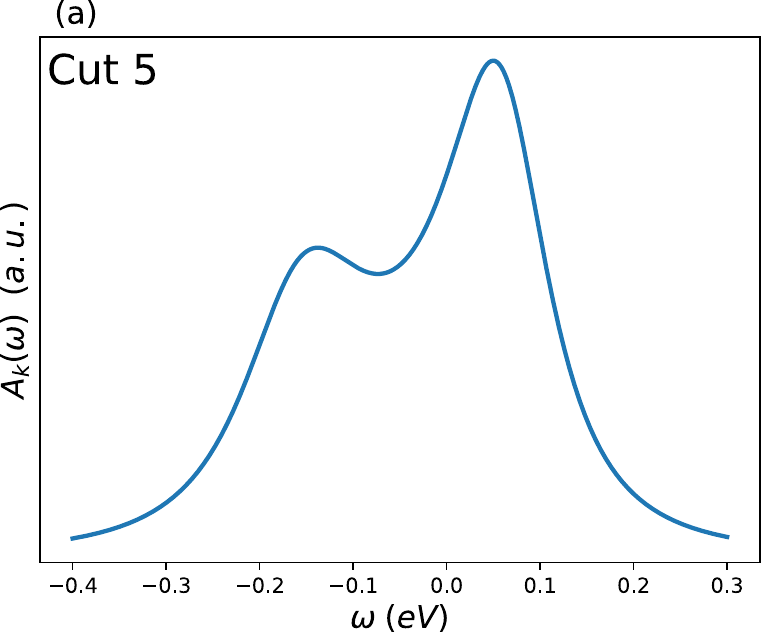}
    \phantomsubcaption\label{fig:15a}
\end{minipage}
\hfill
\begin{minipage}{.45\textwidth}
    \centering
    \includegraphics[width=\linewidth]{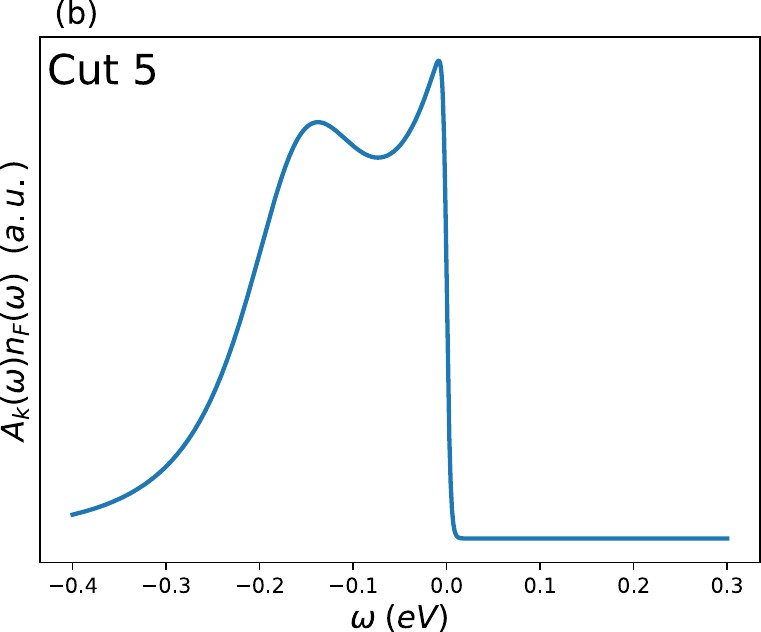}
    \phantomsubcaption\label{fig:15b}
\end{minipage}
\vspace{0cm}
\begin{minipage}{.45\textwidth}
    \centering
    \includegraphics[width=\linewidth]{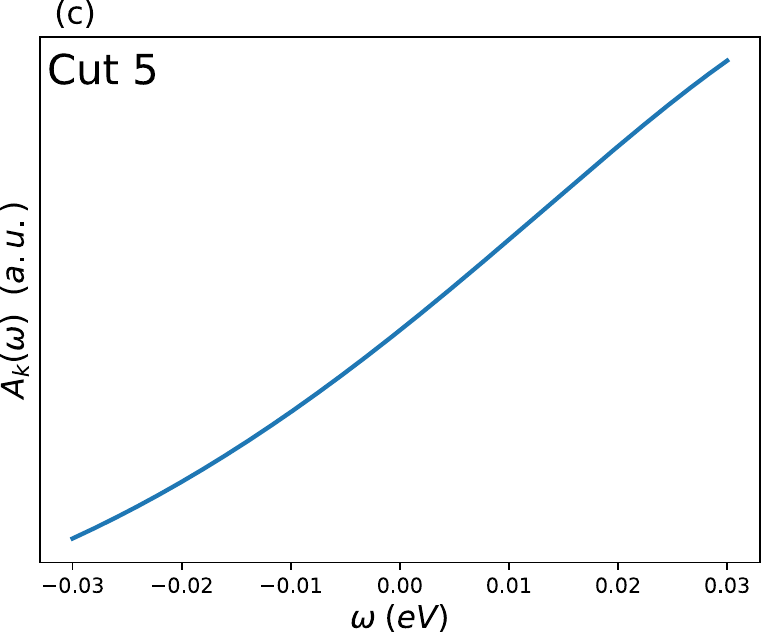}
    \phantomsubcaption\label{fig:15c}
\end{minipage}
\hfill
\begin{minipage}{.45\textwidth}
    \centering
    \includegraphics[width=\linewidth]{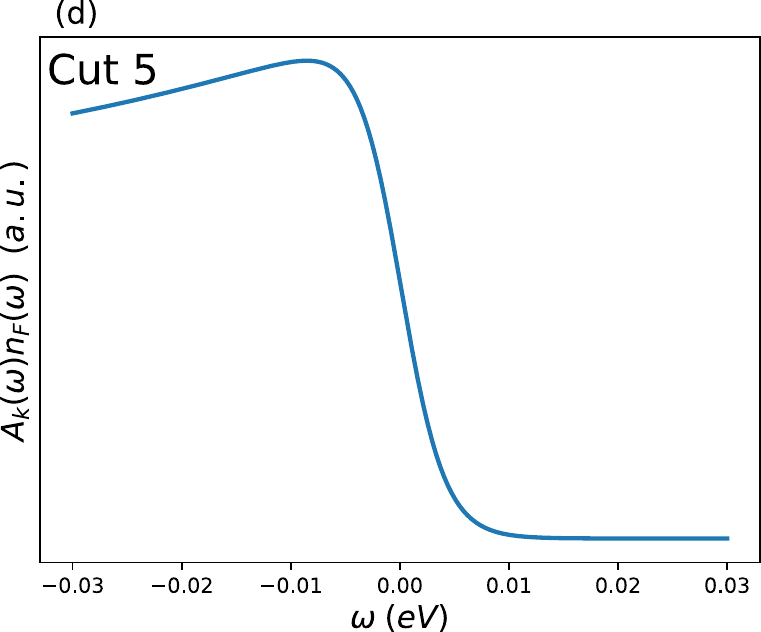}
    \phantomsubcaption\label{fig:15d}
\end{minipage}
\caption{(a)  The spectral function for cut 5 at $k=k_F$ as a function of frequency $\omega$. (b) The spectral intensity (spectral function multiplied by $n_F(\omega)$) for cut 5 at $k=k_F$ as a function of frequency $\omega$. (c)-(d) Spectral function and spectral intensity at $k=k_F$ plotted over a narrower frequency range.}
\label{fig:15}
\end{figure}

We show the low-energy part of the theoretical spectral function and spectral intensity in Figs. \ref{fig:15a} and \ref{fig:15b}, respectively. The second peak in the spectral intensity at small $\omega$ is clearly visible.   For cut 4, there is a plateau at small $\omega$ instead of a peak, and for cuts 1-3, spectral intensity gradually decreases with decreasing frequency.

The presence of the low-energy peak in cut 5, but not in cuts 1-3 can be traced back to the form of the spectral function, Eq. (\ref{2}).  For cut 4, $k_F$ coincides with a hot spot, at which $\epsilon^*_k = \epsilon^*_{k+Q} =0$. Then the spectral function is quadratic in $\omega$ near $\omega =0$, i.e., is flat. Convoluting this spectral function with $n_F (\omega)$ we obtain flat spectral intensity at small negative $\omega$ as in Fig.  \ref{fig:5d}. For other cuts, $\epsilon^*_{k+Q}$ does not vanish at $k_F$, where $\epsilon^*_{k} =0$. We label its finite value at this point as $c$. For cuts 1-3, $c >0$, for cut 5, $c<0$.  Because of finite $c$, the spectral function near $\omega=0$ scales linearly with $\omega$:
\beq
A_{k_F} (\omega) = A_{k_F} (0) \left(1 + Z_c \omega \right)
\eeq
Expanding in (\ref{2}) to linear odder in $\omega$, we obtain
\beq
Z_c = -\frac{1}{1+c^2} \left(c + 2 \lambda_{th} (\lambda_{th} - \sqrt{1+c^2}) \frac{\log{(\sqrt{1+c^2}-c)}}{\frac{\pi^2 \lambda^2_{th}}{4} + \lambda^2_{th} \log^2{(\sqrt{1+c^2}-c)}}\right)
\eeq
{\RaggedRight
At small $c$,}
\beq
Z_c \approx -\frac{4c}{\pi^2\lambda^2_{th}} \left(2 \lambda + \frac{\pi^2-8}{4}\lambda^2\right)
\eeq

We see that the sign of $Z_c$ is opposite to the sign of $c$.  For cut 5, $c<0$, hence $Z_c >0$ and the spectral function {\it increases} as a negative $\omega$ decreases (see Fig. \ref{fig:15c}). Convoluting with the Fermi function, we immediately obtain the peak in the spectral intensity at a negative $\omega \sim -T$ (see Fig. \ref{fig:15d}). For cuts 1-3, $Z_c <0$, i.e.,  the linear term in the spectral function has a negative slope. In this case, the spectral intensity obviously does not possess a second maximum.

\begin{figure}[H]
\begin{center}
\includegraphics[width=.5\textwidth]{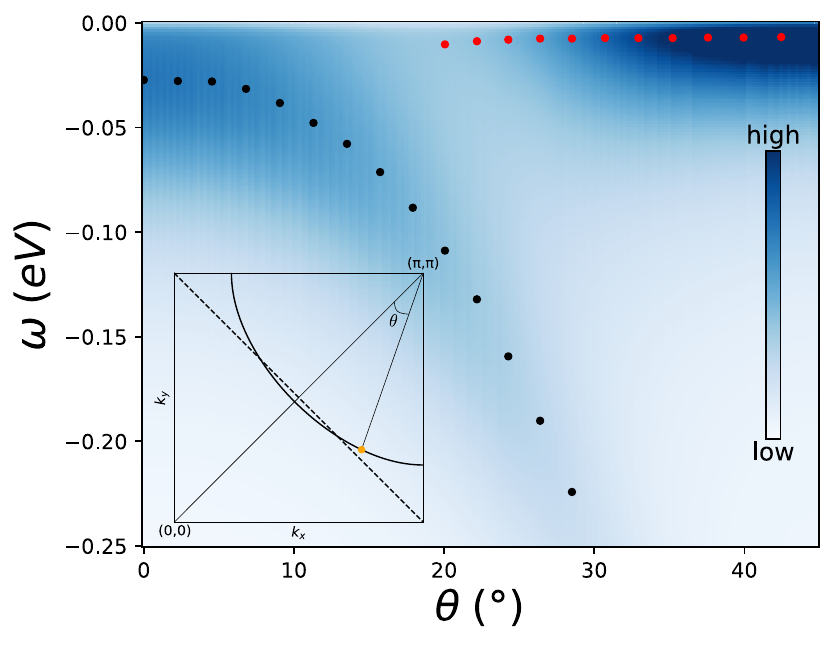}
\end{center}
\caption{Heatmap of the EDC spectra at Fermi momentum as a function of the Fermi surface angle (see inset). The black dots track the position of the EDC peak at large negative frequencies whereas the red dots track the position of the low energy peak.}
\label{fig:16}
\end{figure}

In Fig. \ref{fig:16}, we show  the evolution of spectral weight associated with high-energy and low-energy peaks as one moves along the Fermi surface. We find that the low-energy peak is present for $\theta\gtrsim\theta_{h}$ ($\theta\gtrsim 20^o$)  and stays within a very narrow energy range $\omega\sim7-10\; meV$. This is consistent with the experimental data, presented in Fig. 1E in Ref. \cite{Xu_2024}.

\end{document}